\documentclass[traditabstract]{aa}

\usepackage{graphicx,color}
\usepackage{amssymb,verbatim,hyperref}
\usepackage{txfonts}
\usepackage{natbib}
\usepackage{mathptmx}

\begin{document}

\title{Extreme AGN feedback in the fossil galaxy group SDSSTG 4436} 

\author{D. Eckert\inst{\ref{gva}} \and F. Gastaldello\inst{\ref{iasfmi}} \and L. Lovisari\inst{\ref{iasfmi},\ref{cfa}} \and S. McGee\inst{\ref{ubirm}} \and T. Pasini\inst{\ref{ira}} \and M. Brienza\inst{\ref{ira}} \and K. Kolokythas\inst{\ref{rhodes},\ref{sarao}} \and E. O'Sullivan\inst{\ref{cfa}} \and  A. Simionescu\inst{\ref{sron}} \and M. Sun\inst{\ref{ualab}} \and M. Ayromlou\inst{\ref{bonn}} \and M. A. Bourne\inst{\ref{hertf},\ref{ucamb},\ref{kavli}} \and Y. Chen\inst{\ref{tsing}} \and W. Cui\inst{\ref{umad},\ref{ciaff}} \and S. Ettori\inst{\ref{oabo}} \and A. Finoguenov\inst{\ref{uhel}} \and G. Gozaliasl\inst{\ref{aalto}} \and R. Kale\inst{\ref{pune}} \and F. Mernier\inst{\ref{gsfc},\ref{umar}} \and B. D. Oppenheimer\inst{\ref{ucol}} \and G. Schellenberger\inst{\ref{cfa}} \and R. Seppi\inst{\ref{gva}} \and E. Tempel\inst{\ref{tartu},\ref{tallinn}}}

\institute{
Department of Astronomy, University of Geneva, Ch. d'Ecogia 16, 1290 Versoix, Switzerland\label{gva}\\
\email{Dominique.Eckert@unige.ch}
\and
INAF - IASF Milano, Via Alfonso Corti 12, 20133 Milan, Italy\label{iasfmi}
\and
Center for Astrophysics | Harvard \& Smithsonian, 60 Garden Street, Cambridge, MA 02138, USA\label{cfa}
\and
School of Physics and Astronomy, University of Birmingham, Birmingham, B152TT, UK\label{ubirm}
\and
INAF - Istituto di Radioastronomia, via P. Gobetti 101, 40129, Bologna, Italy\label{ira}
\and
Centre for Radio Astronomy Techniques and Technologies, Department of Physics and Electronics, Rhodes University,\\ P.O. Box 94, Makhanda 6140, South Africa\label{rhodes}
\and
South African Radio Astronomy Observatory, Black River Park North, 2 Fir St, Cape Town, 7925, South Africa\label{sarao}
\and
SRON Netherlands Institute for Space Research, Niels Bohrweg 4, NL-2333 CA Leiden, the Netherlands\label{sron}
\and
Department of Physics and Astronomy, University of Alabama in Huntsville, Huntsville, AL35899, USA\label{ualab}
\and
Argelander Institute für Astronomie, Auf dem Hügel 71, D-53121 Bonn, Germany\label{bonn}
\and
Centre for Astrophysics Research, Department of Physics, Astronomy and Mathematics, University of Hertfordshire, College Lane, Hatfield, AL10 9AB, UK\label{hertf}
\and
Institute of Astronomy, University of Cambridge, Madingley Road, Cambridge CB3 0HA, UK\label{ucamb}
\and
Kavli Institute for Cosmology (KICC), University of Cambridge, Madingley Road, Cambridge CB3 0HA, UK\label{kavli}
\and
Department of Astronomy, Tsinghua University, Beijing 100084, China\label{tsing}
\and
Departamento de Física Teórica, M-8, Universidad Autónoma de Madrid, Cantoblanco 28049, Madrid, Spain\label{umad}
\and
Centro de Investigación Avanzada en Física Fundamental, Universidad Aut\'{o}noma de Madrid, Cantoblanco, 28049 Madrid, Spain\label{ciaff}
\and
INAF – Osservatorio di Astrofisica e Scienza dello Spazio di Bologna, Via P. Gobetti 93/3, 40129 Bologna, Italy\label{oabo}
\and
Department of Physics, University of Helsinki, Gustaf Hällströmin katu 2, 00560 Helsinki, Finland\label{uhel}
\and
Department of Computer Science, Aalto University, PO Box 15400, Espoo, FI-00 076, Finland\label{aalto}
\and
National Centre for Radio Astrophysics, Tata Institute of Fundamental Research, Savitribai Phule Pune University Campus, Ganeshkhind, Pune 411007, India\label{pune}
\and
NASA Goddard Space Flight Center, Code 662, Greenbelt, MD 20771, USA\label{gsfc}
\and
Department of Astronomy, University of Maryland, College Park, MD 20742-2421, USA\label{umar}
\and
CASA, Department of Astrophysical and Planetary Sciences, University of Colorado, 389 UCB, Boulder, CO 80309, USA\label{ucol}
\and
Tartu Observatory University of Tartu, Observatooriumi 1, 61602 Tõravere, Estonia\label{tartu}
\and
Estonian Academy of Sciences, Kohtu 6, 10130 Tallinn, Estonia\label{tallinn}}

\abstract{
Supermassive black hole feedback is the currently favoured mechanism to regulate the star formation rate of galaxies and prevent the formation of ultra-massive galaxies ($M_\star>10^{12}M_\odot$). However, the mechanism through which the outflowing energy is transferred to the surrounding medium strongly varies from one galaxy evolution model to another, such that a unified model for AGN feedback does not currently exist. The hot atmospheres of galaxy groups are highly sensitive laboratories of the feedback process, as the injected black hole energy is comparable to the binding energy of halo gas particles. Here we report multi-wavelength observations of the fossil galaxy group SDSSTG 4436. The hot atmosphere of this system exhibits a highly relaxed morphology centred on the giant elliptical galaxy NGC~3298. The X-ray emission from the system features a compact core ($<$10 kpc) and a steep increase in the entropy and cooling time of the gas, with the cooling time reaching the age of the Universe $\sim15$ kpc from the centre of the galaxy. The observed entropy profile implies a total injected energy of $\sim1.5\times10^{61}$ ergs, which given the high level of relaxation could not have been injected by a recent merging event. Star formation in the central galaxy NGC~3298 is strongly quenched and its stellar population is very old ($\sim$10.6 Gyr). The currently detected radio jets have low power and are confined within the central compact core. All the available evidence implies that this system was affected by giant AGN outbursts which excessively heated the neighbouring gas and prevented the formation of a self-regulated feedback cycle. Our findings imply that AGN outbursts can be energetic enough to unbind gas particles and lead to the disruption of cool cores. }

\keywords{X-rays: galaxies: clusters - Galaxies: clusters: general - Galaxies: groups: general - Galaxies: clusters: intracluster medium - cosmology: large-scale structure}
\maketitle

\section{Introduction}

Accreting supermassive black holes (SMBHs) at the centre of galaxies exhibit outflows in the form of jets and winds that interact with the gaseous medium of their host halo \citep{Laha:2021}. This phenomenon, usually referred as active galactic nucleus (AGN) feedback, is the currently favoured mechanism to solve a number of outstanding problems in galaxy formation \citep[see][for a review]{Fabian:2012}. These include the absence of galaxies with a stellar mass beyond $\sim10^{12}M_\odot$ \citep{Cowie:1996}, the galaxy colour bimodality \citep{Cui2022}, the origin of the relation between SMBH mass and galaxy properties \citep{Kormendy:2013}, the co-evolution between star formation rate and SMBH activity \citep{Madau:1996}, and the over-cooling problem in galaxy cluster cores \citep{McNamara:2007}. In the past decade, AGN feedback has become widely used in galaxy evolution models, to the point that all modern cosmological hydrodynamical simulation suites include a prescription for AGN feedback \citep{Schaye:2015,Weinberger:2018,Dave2019,Tremmel:2017,Henden2018}. The implementation of feedback within these simulations is usually tuned to reproduce the properties of the galaxy populations as closely as possible \citep{Crain:2015}. However, models producing similar galaxy stellar mass functions sometimes make very different predictions on the properties of the hot gaseous haloes that surround galaxies \citep{Oppenheimer:2021} depending on how much feedback energy is injected and how it is deposited within the surrounding medium. 

In this respect, the hot atmospheres of galaxy groups and massive galaxies act as highly sensitive calorimeters of the total energy injected by AGN throughout cosmic time \citep{Eckert:2021,Donahue:2022}. Galaxy groups are usually defined as bound systems of a few tens of galaxies residing within haloes of a total mass in the range $10^{13}-10^{14}M_\odot$ \citep{Lovisari:2021}. Galaxy groups are filled with an intra-group medium (IGrM) with gas temperatures in the range $0.5-2$ keV \citep{Mulchaey:2000}. Compared to their more massive counterparts (galaxy clusters), galaxy groups are usually baryon-poor \citep{Gasta:2007,Sun:2009a,Lovisari:2015,Eckert:2016,Akino:2022,Voit:2024}, which is indicative of a stronger influence of feedback due to their shallower gravitational potential \citep{Gaspari:2014_scalings}. For haloes of a mass of a few $10^{13}M_\odot$, the energy injected by the central SMBH over cosmic time is comparable to the binding energy of gas particles in group cores \citep{Eckert:2021}, such that the properties of the IGrM can be substantially altered by AGN feedback. The scaling relations between IGrM properties and halo mass deviate from expectations from the self-similar model \citep{Kaiser:1986}, including the luminosity-temperature relation \citep{Finoguenov:2006,Maughan:2012,Giles:2016,Lovisari:2021} and the $Y-M$ relation \citep{Yang2022}, which is usually interpreted as evidence for the strong impact of AGN feedback on group cores \citep{McCarthy:2010,LeBrun:2014}. 

The total injected feedback energy in the IGrM is most directly traced by the gas entropy $K=k_B T n_e^{-2/3}$ \citep{Ponman:1999}, which is related to the thermodynamic entropy as $S=k_B\ln K^{3/2}$. In regions where cooling losses are negligible, non-gravitational processes (AGN, supernovae, stellar winds, etc.) can only raise the gas entropy, such that the total non-gravitational energy can be determined by comparing the measured entropy with the baseline gravitational entropy profile expected from structure formation \citep{Voit:2005c}. The required non-gravitational entropy excess is usually found to be larger in low-mass systems than in galaxy clusters \citep{Ponman:1999,Finoguenov:2002,Sun:2009a,Humphrey:2012,Simionescu:2017}. However, owing to the difficulty of selecting representative samples of galaxy groups \citep{Eckert:2011}, the dependence of the total injected non-gravitational energy on halo mass is poorly known, as is the scatter of this relation at fixed halo mass.

In 2022, we were awarded the \emph{XMM-Newton} Group AGN Project \citep{XGAP}, a large programme on the \emph{XMM-Newton} X-ray satellite to observe a sample of galaxy groups selected as bound structures using spectroscopic data from the Sloan Digital Sky Survey (SDSS) using the Friends of Friends (FoF) algorithm \citep{Tempel:2017}. Groups with a minimum of 8 spectroscopically confirmed members were cross-matched with weak, extended X-ray sources discovered in ROSAT all-sky survey data \citep{Damsted:2024} to ensure that the selected systems are virialised and contain an IGrM. From this catalogue, we selected a representative sample of 49 groups for deep X-ray follow-up with \emph{XMM-Newton} \citep{XGAP}, 38 of which had not previously been observed by modern X-ray telescopes. Among the sample is SDSSTG 4436 ($z=0.046$, hereafter S4436), a group containing 31 FoF spectroscopic members and centred on the bright elliptical galaxy NGC~3298 (R.A.=159.301, Dec=+50.120; $z=0.0451$, $m_r=13.6$~mag). The system was clearly detected as an extended source in the ROSAT all-sky survey \citep{Damsted:2024}, with a signal-to-noise of $6.2$ and a $[0.1-2.4]$ keV flux of $(1.60\pm0.26)\times10^{-12}$ ergs s$^{-1}$ cm$^{-2}$. At the redshift of the system, this corresponds to an X-ray luminosity of $(1.2\pm0.2)\times10^{43}$ erg/s. The group exhibits a high magnitude gap between the two brightest members (2.1 magnitude in $r$ band), which classifies the system as a fossil group \citep{Aguerri:2021}. Fossil groups are defined as systems where the magnitude gap between the brightest galaxy and the second brightest member within $0.5R_{200}$ exceeds 2 magnitudes \citep{Jones:2003}. They are thought to be old systems where the dominant galaxy has grown progressively through dry mergers.

Here we present multi-wavelength observations of S4436. We present observations of the IGrM in this system with \emph{XMM-Newton} and study the entropy and cooling time profiles of the gas within the system to determine the impact of AGN feedback on the properties of the surrounding gas over scales of hundreds of kpc. We complement the X-ray data with observations of the central radio galaxy with LOFAR and of the properties of the central galaxy from SDSS MaNGA. At the redshift of $z=0.046$, an angular scale of 1 arcmin corresponds to a physical size of 56 kpc \citep{Planck:2016}, such that the angular resolution of \emph{XMM-Newton} ($\sim15$ arcsec) is sufficient to resolve scales of $\sim14$ kpc.

\section{Data analysis}

\subsection{\emph{XMM-Newton} data analysis}
\label{sec:xmm_analysis}

\subsubsection{Data reduction}

\begin{figure*}[t]
\centerline{\includegraphics[width=0.5\textwidth]{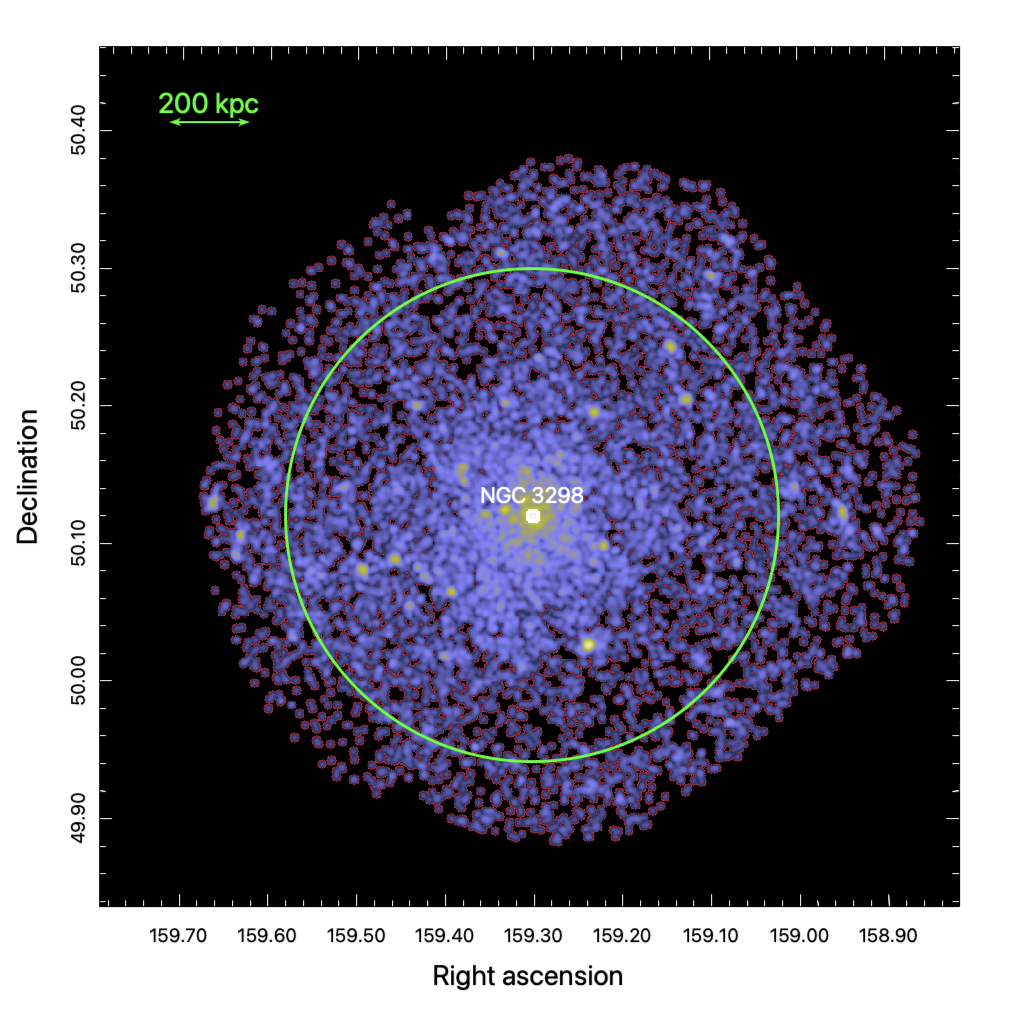}\includegraphics[width=0.5\textwidth]{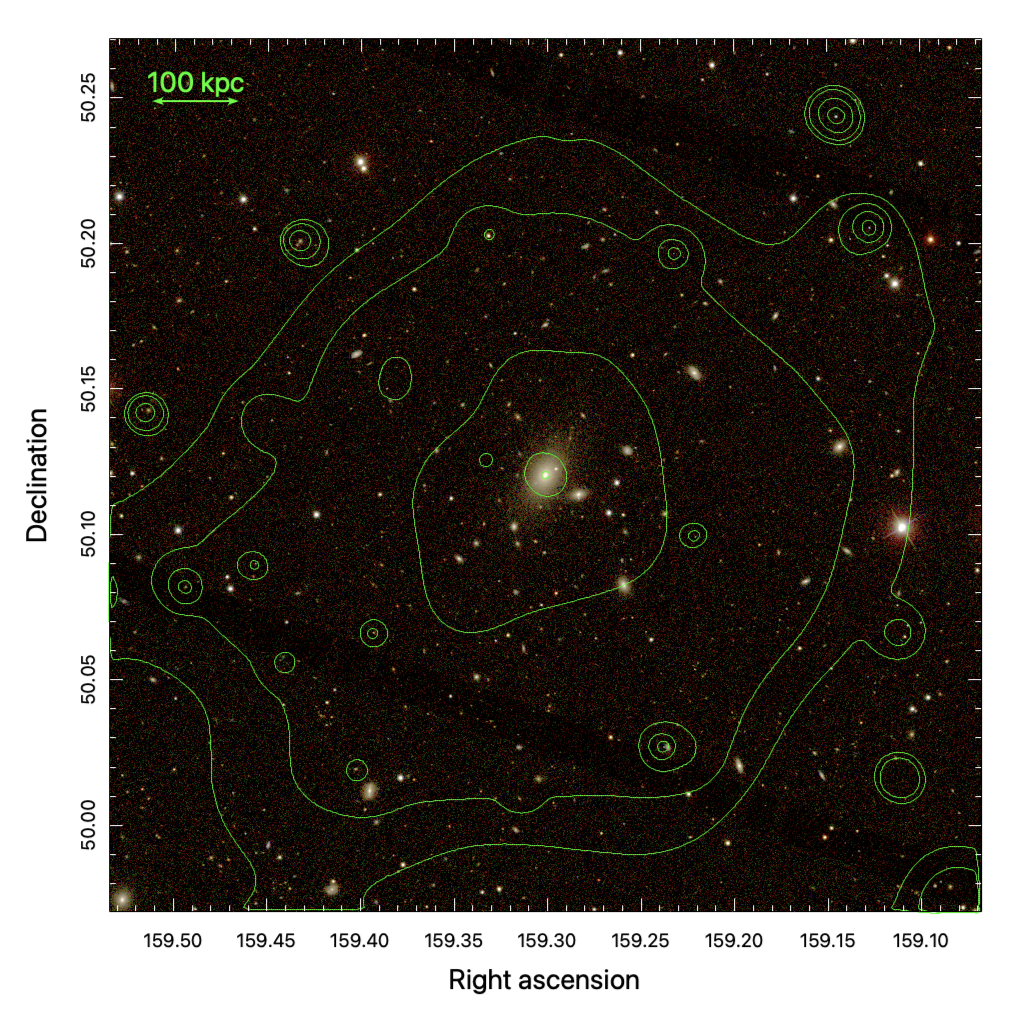}}
\caption{\label{fig:ima}X-ray and optical images of the galaxy group S4436. The left-hand panel shows the \emph{XMM-Newton}/EPIC count map in the [0.7-1.2] keV band, smoothed with a Gaussian kernel of 10 arcsec width. The location of the central galaxy NGC~3298 is indicated with the white square, whereas the green circle shows the approximate location of $R_{500}$. The right-hand panel shows an SDSS RGB map of the system, with R=$i$, G=$r$, and B=$g$. The green contours show X-ray isophotes extracted from the smoothed \emph{XMM-Newton} image.}
\end{figure*}

S4436 was observed by \emph{XMM-Newton} on October 25, 2022 for a total of 22 kiloseconds (ks; $\sim6$h). The \emph{XMM-Newton} observation of S4436 (observation ID 0904700501, PI: D. Eckert) was reduced using the \texttt{XMMSAS} software package, version 19.1, and the X-COP data analysis pipeline \citep{Ghirardini:2019}. After running the standard event screening procedures, we extracted light curves of the observation in the field of view and in the unexposed corners of the three detectors of the European Photon Imaging Camera (EPIC) to filter out time periods affected by flaring background. After filtering out flaring time periods, the total good observing time is 10.9 ks for the EPIC-pn instrument and 16.9 ks for EPIC-MOS. For the two MOS detectors, we excluded the chips operating in anomalous mode (CCD \#4 for MOS1 and CCD \#5 for MOS2). From the clean event lists, we extracted images from all three cameras in the [0.7-1.2] keV band, which optimises the signal-to-background ratio \citep{Ettori:2011}. We used the task \texttt{eexpmap} to extract effective exposure maps for the three detectors including the telescope's vignetting. Maps of the non X-ray background were produced from a large collection of observations in filter-wheel-closed mode, which were then rescaled to match the count rates measured in the corners of the three detectors. The contribution of residual soft protons was estimated from an empirical relation between the difference of high-energy count rates inside and outside the field of view and the normalisation of the soft proton component \citep{Salvetti:2017}. Finally, we produced combined EPIC maps by summing up the individual maps from the three detectors, the exposure maps, and the non X-ray background maps. The resulting total EPIC count map is shown in the left-hand panel of Fig. \ref{fig:ima}. Contaminating point-like sources were detected on the total EPIC count map using the task \texttt{ewavelet} and masked for the extraction of spectral and surface brightness profiles. We also created a map free of point sources by refilling the masked areas using a Poisson realisation of the surrounding background surface brightness. In the right-hand panel of Fig. \ref{fig:ima} we show an SDSS RGB map of the system, with R=$i$, G=$r$, and B=$g$, and X-ray contours superimposed.

\subsubsection{Spectral analysis}
\label{sec:spec}

\begin{figure}
\resizebox{\hsize}{!}{\includegraphics{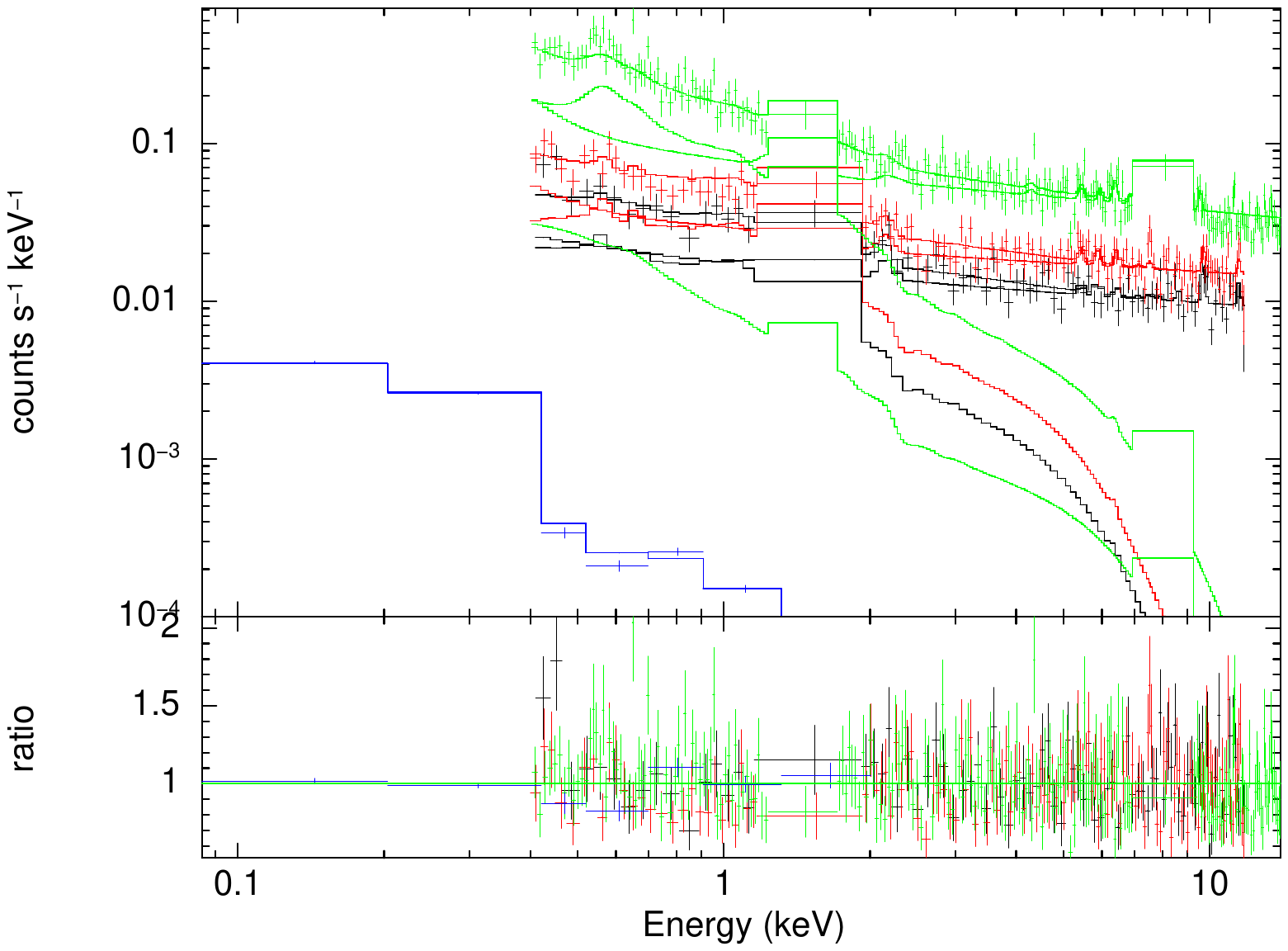}}
\caption{X-ray sky background estimation in the region surrounding S4436. The best fitting three-component model was extracted from \emph{XMM-Newton} EPIC/pn (green), EPIC/MOS1 (black) and EPIC/MOS2 (red) data within an annulus located [12-15] arcmin away from NGC~3298. The data were jointly fitted with the ROSAT all-sky survey data (blue) extracted [1-1.5] degrees away from the core of the group. The bottom panel shows the ratio between the data and the best-fitting joint model.}
\label{fig:bkg}
\end{figure}

The temperature profile of the system was determined by extracting the spectra of 12 logarithmically spaced circular annuli centred on the core of NGC~3298. The spectral analysis technique follows the procedure outlined in \citet{Rossetti:2024}. To determine the contribution of unrelated sky background components within the field of view, we extracted the spectrum from an outermost annulus ([12.5-15] arcmin from the group center) as well as the ROSAT all-sky survey (RASS) spectrum extracted from an annular region located between 1 and 1.5 degrees from NGC~3298\footnote{\href{https://heasarc.gsfc.nasa.gov/cgi-bin/Tools/xraybg/xraybg.pl}{https://heasarc.gsfc.nasa.gov/cgi-bin/Tools/xraybg/xraybg.pl}}. The EPIC and ROSAT spectra were jointly fitted using the \texttt{XSPEC} package with a three-component model including an unabsorbed APEC \citep{Smith:2001} model at a temperature of 0.11 keV for the local hot bubble, an absorbed APEC model with free temperature for the Galactic halo, and an absorbed power law with a photon index of 1.46 for the cosmic X-ray background. We also include a cross-calibration factor of 12\% between the EPIC and RASS spectra \citep{Rossetti:2024}. Our full sky background has four free parameters: the normalisation of the local hot bubble ($N_{LHB}$), the temperature of the Galactic halo ($T_{GH}$) and its normalisation ($N_{GH}$), and the normalisation of the cosmic X-ray background ($N_{CXB}$). The Galactic absorption column density was fixed to the value of $1.13\times10^{20}$ cm$^{-2}$ inferred from the HI4PI survey \citep{HI4PI}. An additional absorbed APEC component was added to the EPIC spectrum alone to allow the possibility of remaining group emission in the background region. To estimate the non X-ray background contribution and spectral shape, we apply the physical background model introduced in \citet{Rossetti:2024}, which accurately predicts the contribution of the cosmic-ray induced component and residual soft protons. The background spectra and the best fitting model are shown in Fig. \ref{fig:bkg}, whereas the best fitting parameters are reported in Table \ref{tab:bkg}. We can see that the model provides an accurate representation of the data, and the procedure results in fairly typical estimates for the celestial X-ray emission \citep[see][]{Rossetti:2024} and the Galactic halo temperature. Group emission on top of the ROSAT background is actually detected in the \emph{XMM-Newton} spectrum of the outermost annulus at the $\sim5\sigma$ level, which shows that X-ray emission from the system is detected out to the edge of the \emph{XMM-Newton} field of view (FoV), which justifies the need for complementing the background spectrum with the RASS data extracted farther away from the system.

\begin{table}
\caption{\label{tab:bkg}Sky background parameters extracted from the joint fitting of the EPIC and RASS background fit (see Fig. \ref{fig:bkg}).}
\begin{center}
\begin{tabular}{cc}
\hline
Parameter & Value \\
\hline
\hline
 $N_{LHB}$ [arcmin$^{-2}$] & $(1.22\pm0.04)\times10^{-6}$ \\
$k_B T_{GH}$ [keV]   & $0.206\pm0.012$ \\
$N_{GH}$ [arcmin$^{-2}$] & $(8.4\pm1.1)\times10^{-7}$ \\
$N_{CXB}$ [arcmin$^{-2}$] & $(8.3\pm0.4)\times10^{-7}$ \\
\hline
\end{tabular}
\end{center}
\end{table}

The total background model extracted from the above approach was then applied to the spectra of all 12 annuli to separate the contribution of the source from that of the X-ray and non X-ray background. The source was modelled as a single-temperature APEC model absorbed by the Galactic column density. The temperature, the metallicity and the normalisation of the APEC model were left free to vary while fitting, whereas the source redshift was fixed to the spectroscopic value of 0.046. The abundance ratios of the various elements were assumed to follow the Solar abundance ratios as defined in the \citet{Asplund:2009} Solar abundance table. To assess the systematic uncertainties associated with the spectral modelling choice, we repeated the analysis by describing the source spectrum with the \texttt{cie} model in the SPEX package \citep{SPEX} and with a Gaussian differential emission measure distribution (\texttt{gadem} model in \texttt{XSPEC}) instead of a single temperature distribution. 

\subsection{SDSS MaNGA data}
\label{sec:sdss}

NGC~3298 was observed by the SDSS MaNGA survey as part of the Massive Nearby Galaxies selection, in which a select number of low redshift ($z < 0.06$) massive galaxies were observed to obtain high spatial resolution.  This galaxy was observed with the 61 fiber configuration which covers the central 22 arcsec ($\sim$ 20 kpc) diameter region. It was observed for a total of 5400 seconds with six 900 second exposures in a 3 point dither pattern to maximize the spatial coverage and resolution. The data reduction pipeline \citep{Law2016} results in  0.5 arcsec$^2$ spaxels across the full region, with a PSF with a FWHM of $\sim$ 2.5 arcsec and a spectral resolution which is $\sim$ 2000 - 2500 and covers 3600 to 10300 angstroms. The data analysis pipeline \citep{Westfall2019} computes stellar continuum and emission line models as well as kinematics and individual line fluxes. Full spectral fitting results were computed using the FIREFLY code \citep{Wilkinson2017} using either the MILES stellar library \citep{Maraston2011, SanchezBlazquez2006} or the MaStar stellar library \citep{Maraston2020, Yan2019}. The MaStar stellar library was specifically designed for MaNGA and allows the full spectral coverage to be constrained and typically results in younger stellar ages and higher metallicities \citep{Neumann2022}. 

\subsection{LOFAR radio data}
\label{sec:radioanalysis}

NGC 3298 lies within the footprint of the Data Release 2 (DR2) of the LOFAR Two-Metre Sky Survey (LoTSS, \citealt{Shimwell:2022}), which provides images of the radio sky at 144 MHz with a resolution of $6^{\prime\prime}$. The source is detected at 144 MHz but appears unresolved, therefore we decided to further process the data by including the LOFAR international stations (IS), which can provide a higher resolution up to 0.3$^{\prime\prime}$. We process the data following the procedure described in \citet{Morabito_2022}, and implemented in the LOFAR-VLBI pipeline\footnote{\url{https://github.com/LOFAR-VLBI/lofar-vlbi-pipeline}}. We summarise here the main steps. 

The LoTSS pointing for which NGC~3298 is closest to the phase centre is P158+50. These survey wide-field images typically have a $\sim 6^{\prime\prime}$ resolution and are produced exploiting the Dutch array of LOFAR (see e.g. \cite{Shimwell:2022}). The inclusion of IS allows instead to push the resolution to $\sim 0.3^{\prime\prime}$. Gain solutions are first derived from the calibrator and applied to the target through the {\ttfamily prefactor} pipeline\footnote{\url{https://github.com/lofar-astron/prefactor}}.  During this step, the data undergoes flagging and averaging and is corrected for polarisation alignment, Faraday rotation, bandpass, clock errors and total electron content (TEC). We then find the best-suited dispersive delay calibrator for our target \citep{Jackson_2016}, which needs to be close and preferentially compact, from the LOFAR Long Baseline Calibrator Survey (LBCS). In the case of NGC 3298, this is ID L333774, located at R.A.=10:34:17.81, Dec=+50:13:29.8. Delay solutions are derived for this calibrator, and applied to the target. The data is then self-calibrated through the {\ttfamily facetselfcal}\footnote{\url{https://github.com/rvweeren/lofar_facet_selfcal}} algorithm.

Imaging is carried out using {\ttfamily WSClean} \citep{Offringa_2014}. Since NGC~3298 has a low surface brightness, it is hardly detected with a $\sim 1^{\prime\prime}$ beam and is not visible at higher resolution. Therefore, we have applied suitable weighting and tapering of the visibilities to obtain a resolution of $\sim 3.5^{\prime\prime}$, where the source is clearly detected. While some calibration artefacts still affect the image, we are able to reach an rms noise of $\sim 200 \mu$ Jy/beam and resolve the lobes of the radio galaxy.

\section{Results}

\begin{figure*}
\centerline{\includegraphics[width=0.5\textwidth]{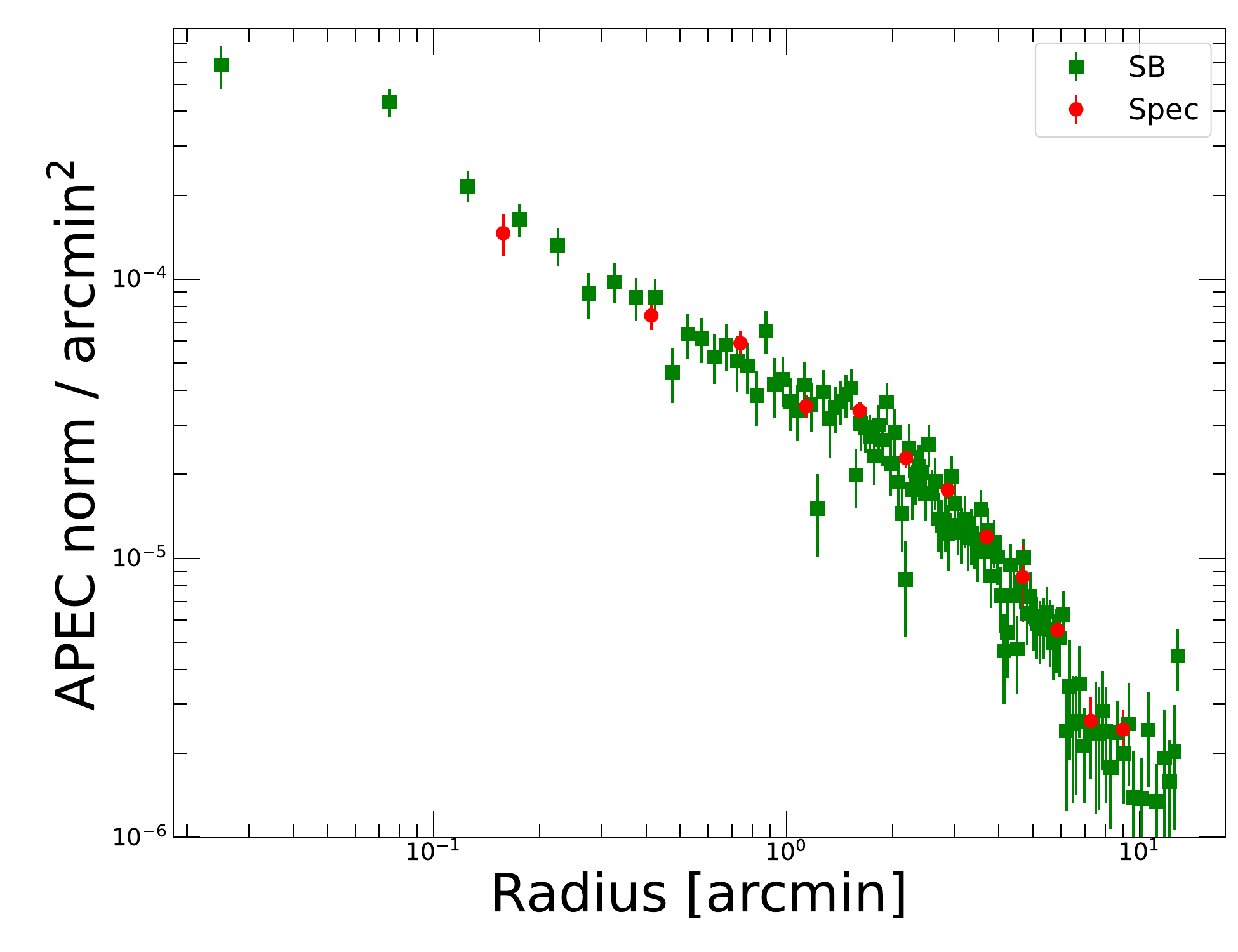}\includegraphics[width=0.5\textwidth]{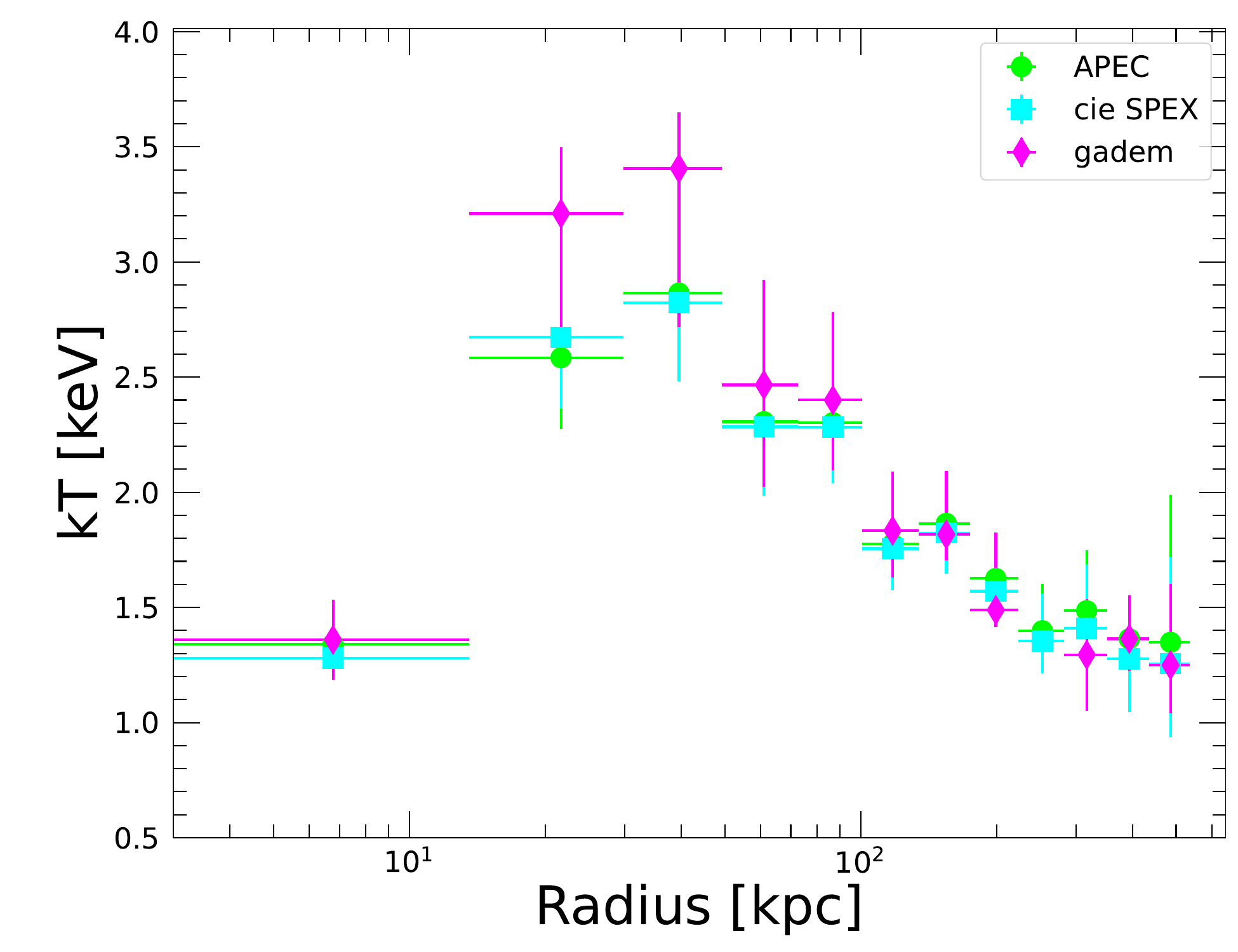}}
\caption{\label{fig:kt_sx} Surface brightness (left) and spectrsocopic temperature profile (right) of S4436. The left-hand panel shows the profile of APEC normalisation per unit area determined directly from the spectral fits (red) and by converting the surface brightness into emission measure using a radially dependent energy conversion factor (green). The right-hand panel shows temperatures retrieved from a single-temperature model using the APEC (green circles) and SPEX (cyan squares) atomic databases, whereas the magenta diamonds represent the temperatures obtained using a Gaussian differential emission measure distribution, in which case the mean temperature of the distribution is displayed. }
\end{figure*}

\subsection{Global properties and dynamical state}

In the left-hand panel of Fig. \ref{fig:ima} we show the combined \emph{XMM-Newton} count map in the [0.7-1.2] keV band, smoothed with a Gaussian kernel of 10 arcsec width (see Sect. \ref{sec:xmm_analysis} for the description of the data reduction scheme). The right-hand panel shows the SDSS $gri$ image of the group with X-ray isophotes overlaid. The X-ray morphology of the group is relaxed, exhibiting approximately circular isophotes centred on the brightest group galaxy (BGG), NGC~3298. The core of the galaxy is associated with a bright, spatially unresolved X-ray source. From the point source free map, we computed the centroid shift $w$ \citep{Mohr:1993}, which determines the variation of the centroid of X-ray emission in decreasing apertures from $R_{500}$ to $0.1R_{500}$. The centroid shift is known to be a good proxy for the dynamical state of the IGrM \citep{Rasia:2013}. For S4436, we measure $w=8.2_{-3.2}^{+5.0}\times10^{-3}$, which firmly classifies the system as dynamically relaxed \citep{Campitiello:2022}. 

The mean temperature of the system within an aperture of 300 kpc is $1.85\pm0.07$ keV, which implies a mass $M_{500}=(7.8\pm1.6)\times10^{13}M_\odot$ and $R_{500}=648$ kpc assuming the weak lensing calibrated mass-temperature relation of \citet{Umetsu:2020}. The retrieved mass is consistent with the value of $(10.7\pm3.5)\times10^{13}M_\odot$ obtained from the velocity dispersion of the member galaxies \citep[$574\pm79$ km/s,][]{Tempel:2017,Damsted:2024} and the $M-\sigma_v$ relation of \citet{Munari:2013}, suggesting that the system is in dynamical equilibrium. Moreover, the large magnitude gap between the dominant galaxy and the second brightest member, which classifies the system as a fossil group. While all fossil groups are not necessarily relaxed, their large central galaxies likely grew through their current size through successive dry mergers \citep[e.g.][]{Lavoie:2016}, which indicates old formation times. Given the relaxed X-ray morphology and the fossil nature of the system, we conclude that the group is dynamically relaxed and has likely not experienced a merger in several Gyr.

\subsection{Temperature and surface brightness profiles}

We extracted the temperature and surface brightness profiles of the system in circular annuli centred on the X-ray peak (see Fig. \ref{fig:kt_sx}). From the [0.7-1.2] keV image of the system, we extracted the surface brightness profile of the system in bins of 3$^{\prime\prime}$ width using the Python package \texttt{pyproffit} \citep{Eckert:2020}. We corrected the surface brightness profile for the telescope's vignetting using the total exposure map and we subtracted the non X-ray background map. The sky background emissivity in the band of interest was computed from the best fitting spectral background model (see Sect. \ref{sec:spec}) and subtracted from the data. To transform the surface brightness profile into an emission measure profile, we fitted the temperature and the emissivity profiles with parametric functions and computed the energy conversion factor at every radius by accounting for variations of temperature and metallicity, which is crucial to properly infer the gas density for plasma in the temperature range 1--2~keV. The normalisation of the APEC model is related to the emission measure as 
\begin{equation}\mbox{Norm} = \frac{10^{-14}}{4\pi(d_A(1+z)^2)}\int n_e n_H \,dV\end{equation}
with $d_A=192$ Mpc the angular diameter distance and $n_e$, $n_H$ the electron and proton number densities, respectively. In the left-hand panel of Fig. \ref{fig:kt_sx} we show the surface brightness profile extracted from the [0.7-1.2] keV image and converted to the APEC normalisation using the radially dependent energy conversion factor. For comparison, the red points indicate the APEC normalisation obtained directly from the spectral fits in circular annuli. We can see that the results of the two approaches are fully consistent, such that we can use the profiles extracted from the surface brightness analysis to extract the gas density profile at higher resolution. 

In the right-hand panel of Fig. \ref{fig:kt_sx} we show the temperature profile of the system extracted from 12 independent radial bins logarithmically spaced from the core to the outskirts (see Sect. \ref{sec:spec}). We also compare the temperatures estimated using the \texttt{APEC} model based on the AtomDB database v3.0.9 \citep{Foster:2010} with the values obtained with the \texttt{SPEX} fitting code v3.07 \citep{SPEX}. We can see that the temperatures measured with the two codes are always consistent, with the \texttt{SPEX} temperatures being on average 3\% lower than the APEC ones. We also checked whether the temperatures obtained under the assumption that the plasma is single-phase within each annulus are consistent with the results obtained assuming a Gaussian differential emission measure distribution (\texttt{gadem}), in which case the width of the distribution was allowed to vary during the fitting procedure together with the mean temperature. We can see in Fig. \ref{fig:kt_sx} that the temperature profile obtained with the Gaussian differential emission measure model agrees well with the results of the single-temperature fit, albeit with substantially larger error bars given the higher complexity of the fitted model. Overall, these tests demonstrate that the results presented here are robust against the choice of the spectral model or atomic database. For the remainder of the paper, we adopt the single-temperature \texttt{APEC} results as our default temperatures.

The surface brightness profile in the innermost regions of the system is very steep, with the surface brightness declining by more than an order of magnitudes in the innermost 1 arcmin (see the left-hand panel of Fig. \ref{fig:kt_sx}). The system  features a bright, compact core coinciding with the central galaxy. The compact core is unresolved by \emph{XMM-Newton}, indicating that its size is less than 10 kpc. The temperature profile of the system shows an abrupt drop in the central regions, from $2.58\pm0.28$ keV at 15 kpc to $1.35\pm0.05$ keV in the innermost 10 kpc. A closer look at the spectrum of the innermost region indicates the presence of a prominent Fe-L emission feature around 1 keV, which provides unambiguous evidence that the spectrum of the compact core is dominated by hot gas and that any contribution of a central point-like source or of a population of unresolved X-ray binaries is negligible. Given that the compact core is spatially unresolved, our data actually provide a lower limit to the temperature drop and the true temperature gradient is likely to be even steeper. 

\subsection{Metallicity profile}

\begin{figure}
\resizebox{\hsize}{!}{\includegraphics{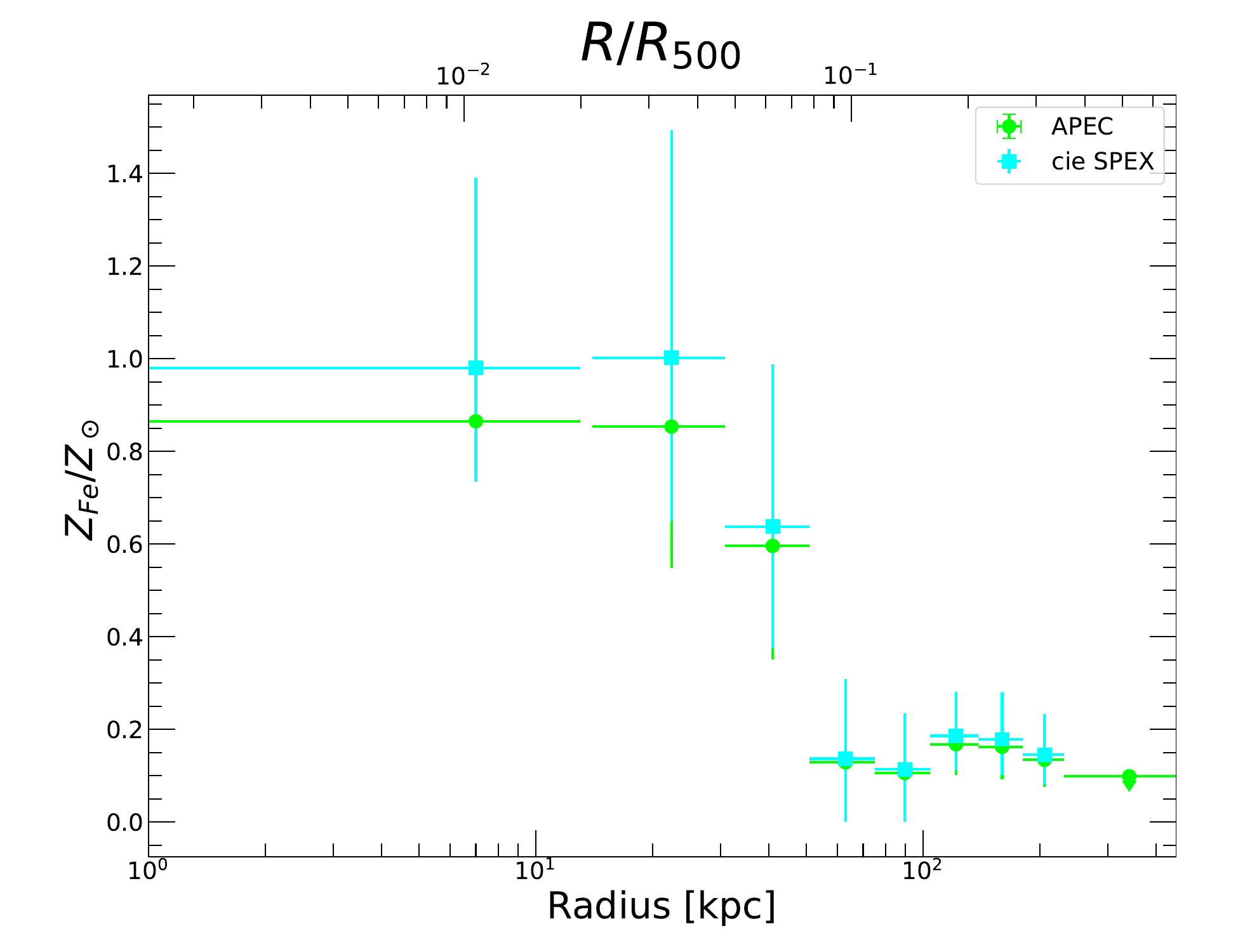}}
\caption{\label{fig:zfe} Metal abundance profile of S4436 as a fraction of the Solar value. The data points show the results of single-temperature fits to the \emph{XMM-Newton} spectra with the APEC (green circles) and SPEX (cyan squares) plasma emission codes. The outermost point is an upper limit to the single-temperature metallicity.}
\end{figure}

We studied the metallicity profile of the system extracted from single-temperature fits to the \emph{XMM-Newton} spectra. To this end, we fix the abundance ratios of every element to the Solar abundance ratios from \citet{Asplund:2009}, and fit a single metallicity value as a ratio of the Solar metallicity. Given the temperature range considered, the constraints mainly arise from the Fe-L complex around 1 keV. We also compared the results obtained with the APEC and SPEX codes. The resulting metal abundance profiles are shown in Fig. \ref{fig:zfe}. We find that the metallicity of the gas is nearly solar within the core ($<40$~kpc) and steeply decreases to $\sim0.1-0.2\,Z_\odot$ beyond 50~kpc. The high-metallicity region extends out to $3-4$ times the radius of the compact core, which shows that the region immediately surrounding the central galaxy has been enriched in metals by supernovae and stellar winds. Beyond this point, the low metallicity of the gas shows that the large-scale halo has not been significantly metal enriched and the metals have been injected prior to the formation of the group \citep[e.g.][]{Werner:2013b}. 

\subsection{Deprojected profiles}
\label{sec:deproj}

\begin{figure*}
\centerline{\includegraphics[width=0.333\textwidth]{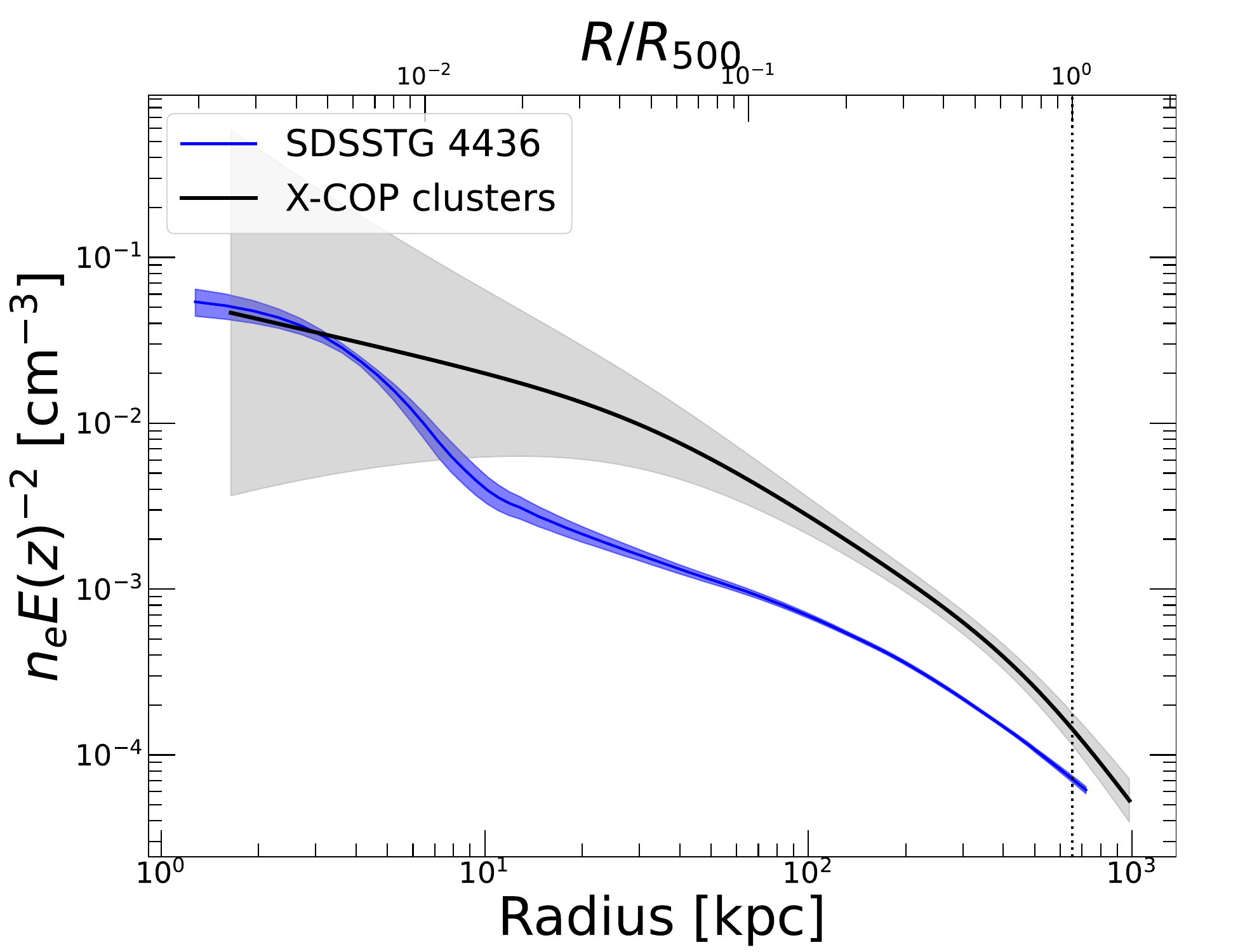}\includegraphics[width=0.333\textwidth]{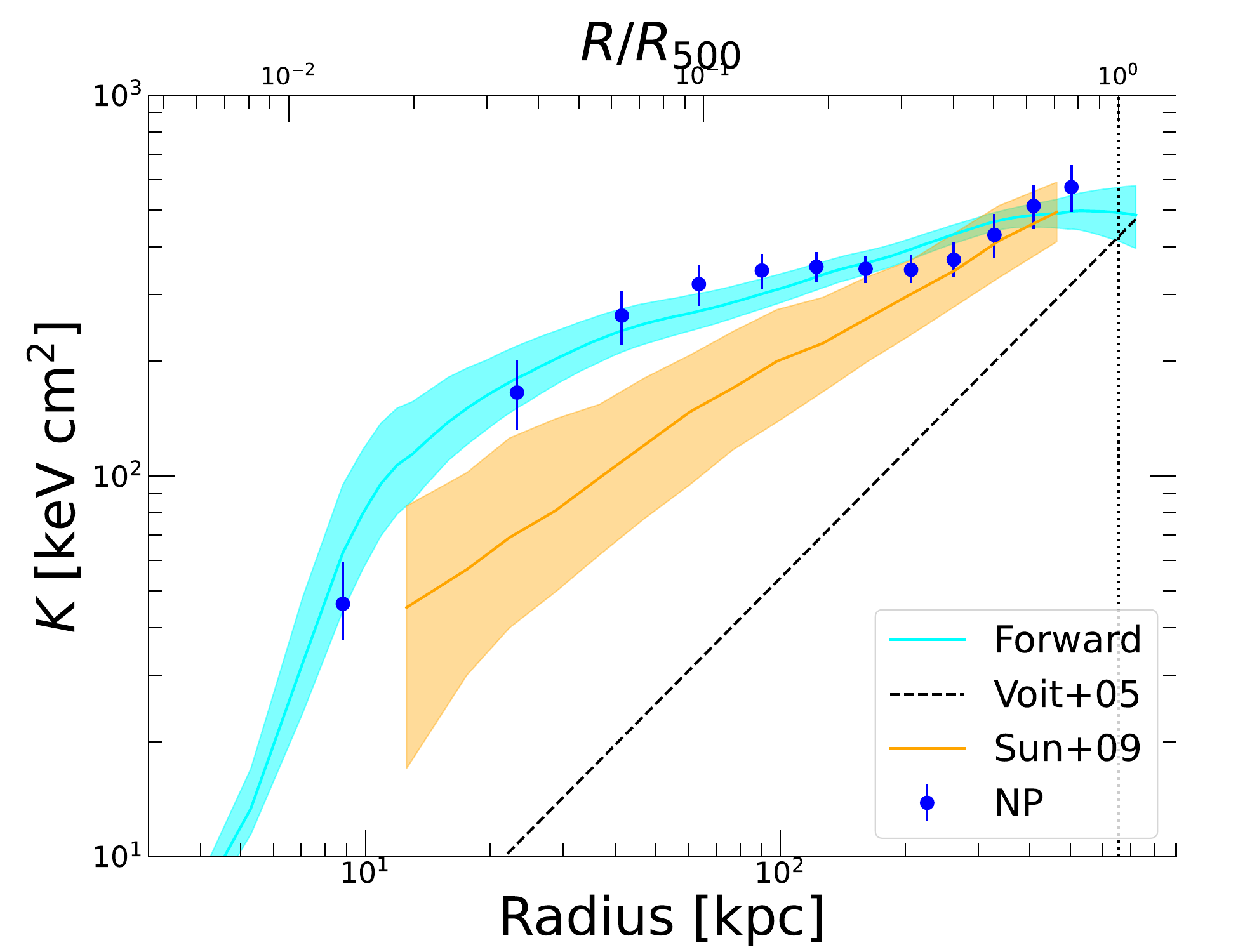}\includegraphics[width=0.333\textwidth]{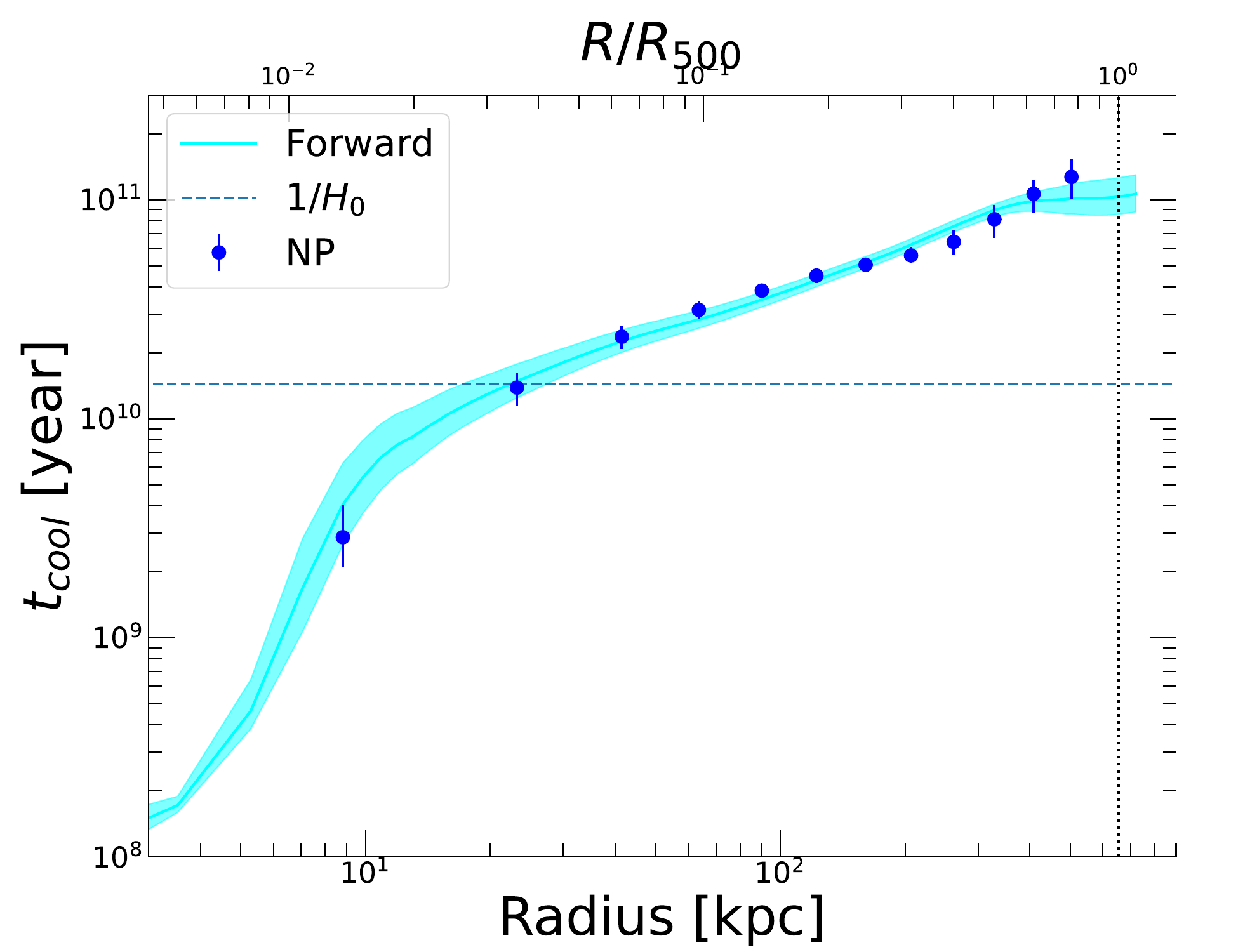}}
\caption{\label{fig:3dprofs}Three-dimensional thermodynamic profiles of the IGrM of S4436. The left-hand panel shows the electron density profile (blue curve). For comparison, the black curve and shaded area show the mean and scatter of the gas density profiles in a sample of massive galaxy clusters \citep{Ghirardini:2019}. The middle panel shows the gas entropy obtained using the parametric (cyan curve) and non-parametric (blue points) deprojection techniques, compared with the entropy profile expected from gravitational collapse \citep{Voit:2005c} (dashed curve) and with the entropy profiles of galaxy groups in the same mass range \citep{Sun:2009a} (orange curve and shaded area). The right-hand panel shows the reconstruction of the gas cooling time, with the approximate age of the Universe, $1/H_0$, indicated by the horizontal dashed line. In all three panels, the dashed vertical line shows the location of $R_{500}=648$ kpc.}
\end{figure*}

To study the three-dimensional profiles of the thermodynamic quantities in the system, we deprojected the observed surface brightness and temperature profiles assuming that the system is spherically symmetric. We used the Python package \texttt{hydromass} \citep{Eckert:2022} to deproject the observed temperature and emission measure profiles and determine the three-dimensional profiles of the various thermodynamic quantities. The gas density profile is modelled using a multi-scale decomposition technique whereby the three-dimensional gas emissivity is described as a linear combination of a large number of basis functions \citep{Eckert:2020}. The basis functions are individually projected, convolved with the \emph{XMM-Newton} PSF, and multiplied by the energy conversion factor profile to compute a model surface brightness profile which is then fitted to the observed surface brightness profile. To assess the systematic uncertainties associated with the temperature deprojection, we consider two different deprojection methods and present the results of both techniques in Fig. \ref{fig:3dprofs}. We either parameterise the 3D electron pressure profile with a  generalised Navarro-Frank-White (gNFW) profile \citep[hereafter Forward]{Nagai:2007} or apply a non-parametric deprojection (hereafter NP) whereby the 3D temperature profile is described as a linear combination of log-normal functions. In both cases, the 3D model is projected along the line of sight to predict the projected temperature profile. More details on the deprojection and PSF deconvolution techniques are provided in Appendix \ref{app:deproj}. 

The deprojected entropy profile is obtained by combining the posterior distributions of the model gas density and temperature. The associated uncertainties are calculated as the 16th and 84th percentiles of the posterior envelopes at all radii. Overall, the 3D entropy profiles obtained with the two techniques are very similar, such that the modelling uncertainties do not affect the results presented here. From the deprojected profiles, we also estimate the gas cooling time, which is defined as the thermal energy of gas particles divided by their cooling rate,
\begin{equation} t_\mathrm{cool}=\frac{3/2 n_{\rm gas} k_B T}{n_e n_H \Lambda(T,Z)}, \end{equation}
with $n_{\rm gas}=n_e+n_i$ and $\Lambda(T,Z)$ the bolometric cooling function, which is a function of gas temperature and metallicity. To recover the cooling time profile, at each radius we use the model temperature and metallicity to calculate the bolometric cooling function using the APEC model, in the same way as for the calculation of the energy conversion factor. We then combined the model density and temperature profiles with the recovered cooling function to compute the posterior distribution of cooling times. 

The resulting profiles of 3D electron density, entropy, and cooling time are shown in Fig. \ref{fig:3dprofs}. The gas density profile of the system shows a two-component behaviour, with a steeply declining profile in the innermost 10 kpc ($\sim0.02\,R_{500}$) and a flat, low-density component beyond $\sim20$ kpc. Such a behaviour is very different from what we typically find in galaxy clusters. For comparison, we show the mean and scatter of the gas density profiles of galaxy clusters in the mass range $4-10\times10^{14}M_\odot$ \citep{Ghirardini:2019}. We can see that beyond the central compact core, S4436 is highly evacuated, with a gas density that lies about an order of magnitude below that of massive clusters at $0.05\,R_{500}$. The behaviour of the density profile is reflected in the gas entropy. In a stratified gaseous atmosphere, we expect the entropy to follow a simple radially increasing behaviour, with the low-entropy gas lying at the bottom of the potential well \citep{Voit:2005c}. We observe that the entropy of S4436 rises very steeply from the centre until it reaches a value of $\sim200$~keV\,cm$^2$ at 20~kpc. The dashed line in Fig. \ref{fig:3dprofs} (center) shows the self-similar entropy profile, which can be described as \citep{Pratt:2010}
\begin{equation}K_{SSC}(R) = 1.42 K_{500} \left(\frac{R}{R_{500}}\right)^{1.1}\label{eq:K_baseline}\end{equation}
with the self-similar normalisation $K_{500}$ given by
\begin{equation}K_{500}=106 \left( \frac{M_{500}}{10^{14} M_\odot}\right)^{2/3} f_b^{-1} E(z)^{-2/3} [\mbox{keV cm}^2] \end{equation}
with $f_b\sim0.15$ the cosmic baryon fraction and $E(z)=[\Omega_m(1+z)^3+\Omega_\Lambda]^{1/2}$.  For S4436 ($z=0.046, M_{500}=7.8\times10^{13}M_\odot$), $K_{500}=298$ keV cm$^2$. We can see that the measured entropy at 20 kpc exceeds the gravitational collapse expectation by more than an order of magnitude. For comparison, the orange curve and shaded area show the range of entropy profiles for galaxy groups in the same mass range from the archival \emph{Chandra} study of \citet{Sun:2009a}. The entropy profile of S4436 occupies the upper end of the range of values in the \citet{Sun:2009a} study, which indicates a very high injection of non-gravitational energy within the group's core. In the right-hand panel of Fig. \ref{fig:3dprofs} we show the profile of the gas cooling time. We can see that the cooling time is short in the very central regions ($t_\mathrm{cool}<1$ Gyr), but it rises very steeply with radius and reaches the age of the Universe at $\sim15$ kpc, i.e. in the outskirts of NGC~3298. Therefore, beyond the central compact core the gas is not efficiently cooling and there is no supply of external gas to the central galaxy. 

\subsection{Radio galaxy}

To identify the source of non-gravitational energy in the system, we gathered the existing radio observations from publicly available surveys. Images from LoTSS \citep{Shimwell:2022} at 6 arcsec resolution reveal the existence of a compact radio source coinciding with NGC~3298, mildly extended from North-West to South-East. The flux density of the source as measured within the 3$\sigma$ contours is $S_{144\mathrm{MHz}}=26.6\pm 4$ mJy, corresponding to a power of $\sim1.4\times10^{23}$ W/Hz at the galaxy's redshift. In order to constrain the morphology of the source at higher resolution, we have calibrated data from the corresponding pointing of LoTSS to include the LOFAR IS across Europe, following the procedure described in Sect.~\ref{sec:radioanalysis}. This allowed us to reach a resolution of $\sim 3.5^{\prime\prime}$ and clearly resolve two lobes symmetrically departing from the galaxy with a total extension of $\sim$10 kpc (see Fig. \ref{fig:lofar_manga}), thus clearly confined within the central compact core. The source is also visible at 1400 MHz in the Faint Images of the Radio Sky at Twenty-Centimeters \citep[FIRST,][]{Becker:1995} at 5 arcsec resolution with a flux density of $S_{1400\mathrm{MHz}}=4.9 \pm 0.5$ mJy, corresponding to a power of $\sim3\times10^{22}$ W/Hz. The measured integrated spectral index $\rm \alpha_{1400}^{144}=0.74 \pm 0.08$ is consistent with freshly accelerated plasma from active jets. Overall, the compact size and the very low radio power of the source indicate that the central AGN is not currently injecting energy in the large-scale group halo. Moreover, the absence of any radio emission beyond the innermost $\sim10$ kpc suggests that any previous large-scale jet activity must have been quenched no less than $\sim100$ Myr ago, which is the typical cooling timescale for relativistic electrons in $\mu$G-level magnetic fields. The oldest known radio-detected feedback features in a galaxy group are $\sim350$ Myr old \citep{Brienza:2021}, which sets a strict lower limit of a few hundred Myr to the epoch of the last feedback episode.

\subsection{Stellar population synthesis of NGC 3298}

\begin{figure*}
\centerline{\includegraphics[width=0.4\textwidth]{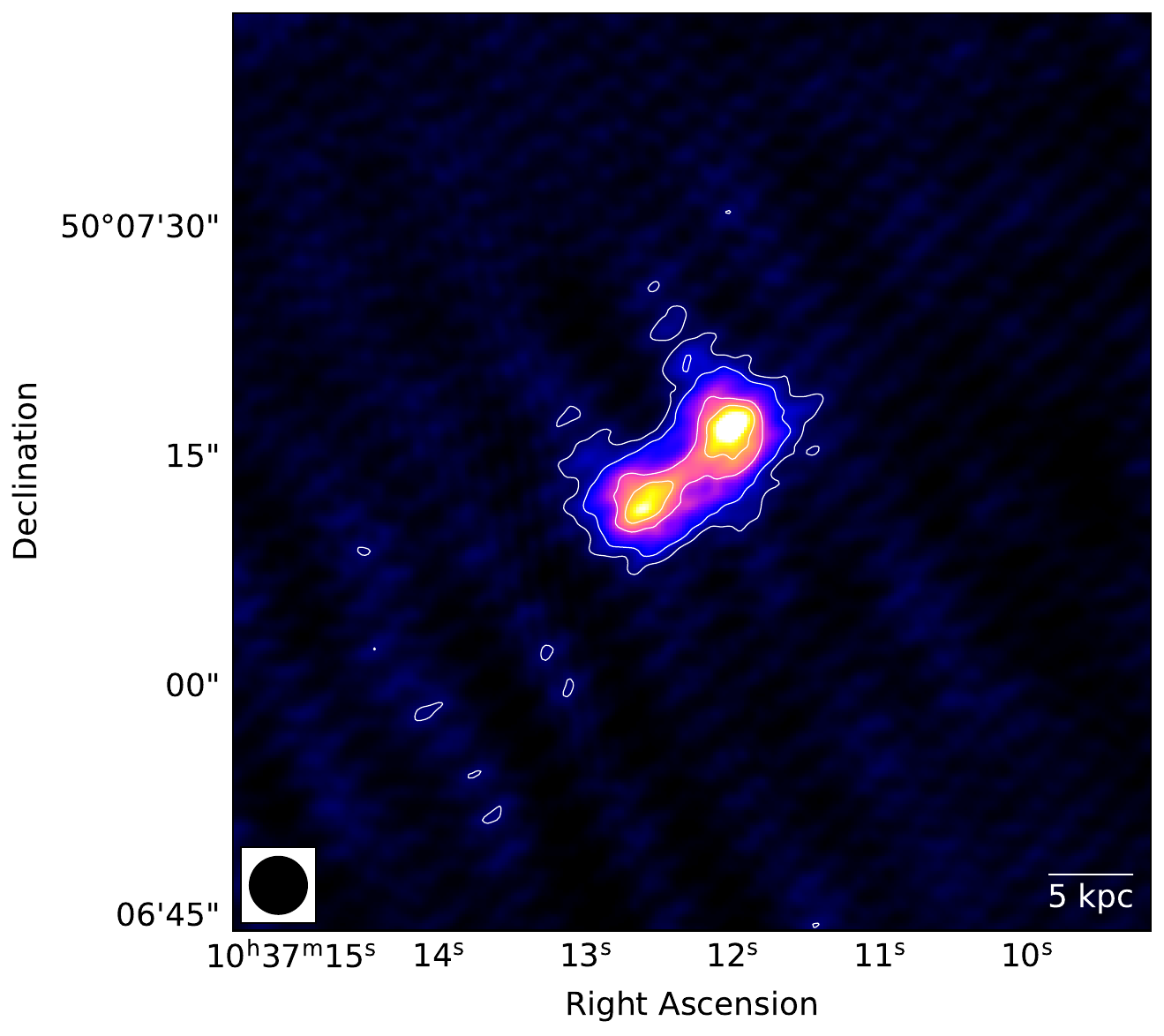}\raisebox{0.5cm}{\includegraphics[width=0.65\textwidth]{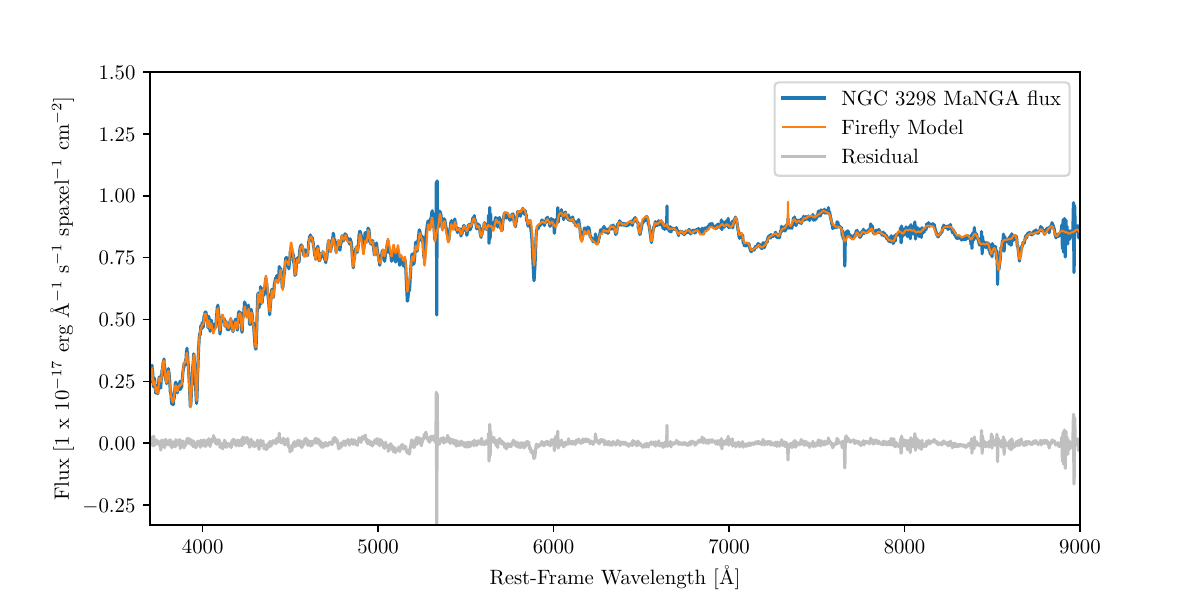}}}
\caption{\label{fig:lofar_manga} \textit{Left}: LOFAR 144 MHz image of NGC~3298 produced using International Stations (IS). The beam (shown in the inset) is $3.5^{\prime\prime} \times 3.5^{\prime\prime}$, and the \textit{rms} noise is $\sim$200 $\mu$Jy beam$^{-1}$. Contours are at 3,6,12,24,48 $\times$ \textit{rms}. \textit{Right}: SDSS MaNGA IFU flux data of NGC~3298. The observed flux (blue line) is the mean of all spaxels with S/N $>$ 15 and is shown with the best fit Firefly model (orange line) with the MaStar stellar library (see Sect. \ref{sec:sdss}). The residuals (grey line) show only minor, non-systematic differences.   
}
\end{figure*}

We also analysed the stellar populations of the central host galaxy using data from the MaNGA integral field unit on the SDSS telescope (see Sect. \ref{sec:sdss}). We fitted the spectrum of NGC~3298 with the FIREFLY code (see the right-hand panel of Fig. \ref{fig:lofar_manga}). The results obtained with the MaStar and MILES methods are largely consistent. In the central 3 arcsec, the MILES method gives a mass-weighted age of 10.69 $\pm$ 1.07 Gyrs and MaStar gives 10.58 $\pm$ 1.08 Gyrs. Interestingly, the MILES method gives a mass weighted age gradient with radius that is consistent with zero (0.020 $\pm$ 0.025 dex / R$_e$), while MaStar has a negative gradient (-0.171 $\pm$ 0.034 dex / R$_e$). Similarly, while the central mass-weighted stellar metallicity is similar, with MILES giving a slightly higher metalllcity (Z = 0.296 $\pm$ 0.040) then MaStar (Z = 0.233 $\pm$ 0.018). The MILES method gives a somewhat negative radial gradient (-0.047 $\pm$ 0.024), whereas MaStar is consistent with zero (0.009 $\pm$ 0.015). In any case, while there are some variations in the age and metallicity between methods, they both agree that this is a very old stellar population with a high metallicity. There is no indication of emission lines indicative of star formation or AGN activity, which classifies the radio source as a low-excitation radio galaxy \citep{Buttiglione:2010}. The average equivalent width of H$\alpha$ in the spaxels is 0.071 $\pm$ 0.211 angstroms, which is consistent with no star formation. The galaxy has a high stellar mass ($\log M_\star/M_\odot = 11.5$) and a very low star formation rate, with no detection of H$\alpha$ emission (see Fig. \ref{fig:lofar_manga} and Sect. \ref{sec:sdss}). The galaxy features a high velocity dispersion of 300 km/s over the entire MaNGA field of view. The stellar component dominates the mass budget within the innermost $\sim10$ kpc, corresponding to the size of the compact core. The high stellar age implies that the galaxy has quenched a very long time ago \citep{Thomas:2010}.

\section{Discussion}

\subsection{Origin of the high-entropy core}

As highlighted in Fig. \ref{fig:3dprofs}, the entropy of S4436 substantially exceeds the typical value expected for galaxy groups of similar mass. The profiles of entropy and cooling time in galaxy groups and clusters can be typically described by a power law with a central floor \citep{Cavagnolo:2009} or a broken power law \citep{Panagoulia:2014_prof,Babyk:2018b}, with the entropy slope gradually steepening with radius from an inner slope of $\sim2/3$ to the self-similar slope of $1.1$ \citep{Babyk:2018b}. The entropy excess is localised within the central regions \citep{Sun:2012} and at large radii the slope and normalisation correspond with the gravitational collapse expectation \citep{Pratt:2010,Ghirardini:2019}. In the case of S4436, the entropy exhibits a very steep central slope ($1.5-2$ in the innermost 10 kpc) and becomes very flat beyond this point ($d\ln K/d\ln R\sim0.3$). This behaviour implies that an unusually large amount of energy was injected into the system, which was responsible for evacuating the core of the group almost completely. Beyond the compact core, the entropy and the cooling time are so large that the gas never cooled down since the period when the energy injection occurred and the system could not form an extended cool core. The large magnitude gap between the BGG and the second brightest member, and the relaxed X-ray morphology, rule out recent merging events as a potential source of energy, such that the observed entropy excess must be of non-gravitational origin. 

The discovery of S4436 reveals the existence of a class of high-entropy systems whereby the injection of non-gravitational energy has prevented the formation of a classical cool core and rapidly quenched the star formation activity in the central galaxy by exhausting the supply of fresh gas from the surrounding halo. Other systems with qualitatively similar features are ESO 3060170 \citep{Sun:2004}, AWM 5 \citep{Baldi:2009b}, and AWM 4 \citep{OSullivan:2010,Gasta:2008}. These systems are all relaxed systems of similar mass with a heated core, although none of them appears as extreme as S4436. ESO 3060170 appears to be a close analog, as it is another fossil group that features a steep increase in its entropy and cooling time profiles within its innermost regions. However, its entropy at 20 kpc from the centre is less than a third of the value reported here for S4436. On the other hand, AWM 4 and AWM 5 both host moderately powerful radio galaxies ($P_{1.4GHz}>5\times10^{23}$ [W/Hz]) with radio lobes extending to extragalactic scales, which provides clear evidence for ongoing heating. In particular, the BGG of AWM 4 is associated with the powerful radio galaxy 4C +24.36 \citep[$\log P_{1.4GHz} = 24.15$,][]{Giacintucci:2008}, which extends over $\sim75$ kpc. This is in stark contrast with the case of S4436, where the current low-power radio jets (see Fig. \ref{fig:lofar_manga}) are currently not injecting much energy into the surrounding IGrM. This implies that previous episodes of AGN activity were able to heat the IGrM to such high levels that the gas located beyond the compact core does not efficiently cool, which has prevented the formation of a new cool core following the end of the last feedback episode. This interpretation is supported by the metallicity profile of the source (Fig. \ref{fig:zfe}), which drops sharply beyond $\sim50$ kpc, implying that metal enrichment at late times from stellar mass loss was not redistributed beyond the core. Therefore, most of the injected non-gravitational energy may have been injected early on in the formation path of the group, as indicated by \citet{Heckman:2024}, who showed that half of the cumulative jet power in massive galaxies is injected at redshifts 1-2.

\subsection{Energy budget calculation}
\label{sec:energy}

The total non-gravitational energy budget in the IGrM can be calculated by contrasting the observed entropy profile with the expected entropy profile from gravitational collapse. Numerical simulations \citep{Voit:2005c,Borgani:2005} predict that gravitational processes lead to a stratified atmosphere where the low-entropy gas sinks to the bottom of the potential well, whereas the high-entropy gas expands and fills larger volumes. Neglecting cooling losses, the difference between the measured entropy and the baseline (Eq. \ref{eq:K_baseline}) can be used to estimate the excess heat injected by non-gravitational processes. The heat element $dQ$ is given by
\begin{equation}dQ = TdS = k_B T \frac{dK}{K}.\end{equation} 
In an isochoric process (no change in volume), the excess heat per particle becomes \citep{Finoguenov:2008,Chaudhuri:2012}
\begin{equation}\Delta Q \approx \frac{k_BT}{\gamma - 1}\frac{K_\mathrm{obs}-K_\mathrm{SSC}}{K_\mathrm{obs}}\label{eq:dQ}.\end{equation}
We note that the isochoric approximation is not strictly valid as a fraction of the gas is ejected from the halo. The total heat accounting for expansion of the volume would be slightly larger than the above estimate, such that the energy estimate presented here provides a lower limit to the true non-gravitational energy. The total non-gravitational energy within radius $R$ is obtained by integrating Eq. \ref{eq:dQ} over all the gas particles,
\begin{equation}E_{NG}(<R) =  \int_0^R \frac{k_BT}{(\gamma-1)\mu m_p} \frac{K_\mathrm{obs}-K_\mathrm{SSC}}{K_\mathrm{obs}} 4\pi r^2 \rho_\mathrm{gas}\, dr . \label{eq:Eng}\end{equation}
Integrating Eq. \ref{eq:Eng} out to $R_{500}$, we estimate a total non-gravitational energy of $\sim1.5\times10^{61}$ erg within $R_{500}$. 

\begin{figure}
\centerline{\resizebox{\hsize}{!}{\includegraphics{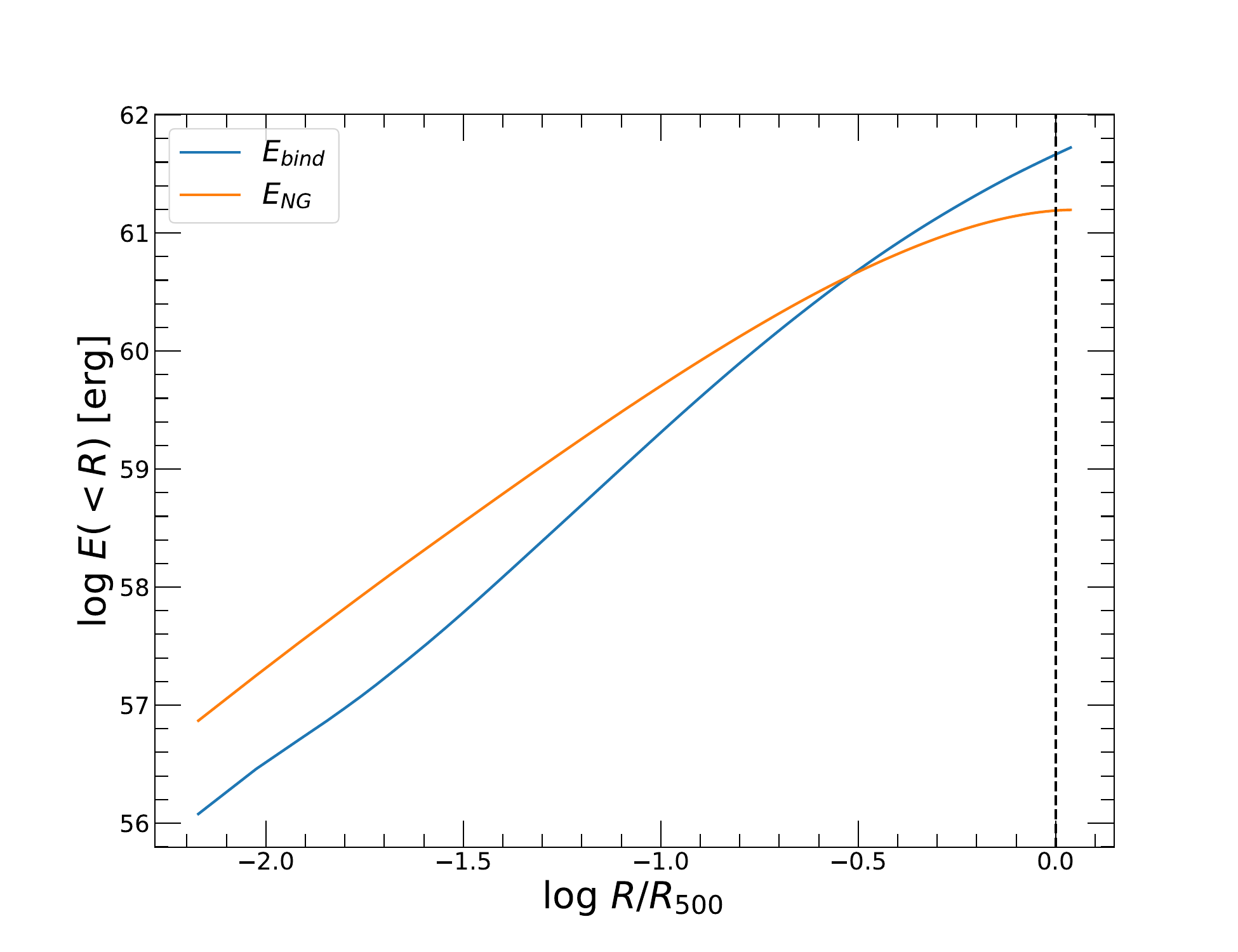}}}
\caption{\label{fig:ebind_eagn}Energy budget of the IGrM of S4436. The blue curve shows the gas binding energy profile obtained through Eq. \ref{eq:Ebind}, whereas the orange curve shows the integrated non-gravitational energy profile from Eq. \ref{eq:Eng}. }
\end{figure}

It is interesting to contrast the above estimate against the binding energy of the gas to study the impact of feedback on the IGrM. The potential energy of a gas mass element $dM_\mathrm{gas}$ at radius $R$ is given by 
\begin{equation}d\Omega = -\frac{GM(<R)}{R}dM_\mathrm{gas}\end{equation}
such that the total binding energy within radius $R$ becomes
\begin{equation}E_\mathrm{bind}(<R)=-G\int_0^R \frac{M(<r)}{r}4\pi r^2 \rho_\mathrm{gas} \, dr . \label{eq:Ebind}\end{equation}
We assume that the total mass profile can be described by a Navarro-Frenk-White model \citep{Navarro:1996} with $M_{500}=7.8\times10^{13}M_\odot$ and a concentration $c_{500}=4$, which is typical of massive groups \citep[e.g.][]{Duffy:2008} and provides a good match to the hydrostatic mass profile of S4436 (see Appendix \ref{app:deproj}). Inserting this model into Eq. \ref{eq:Ebind} returns a total binding energy of $\sim4\times10^{61}$ erg within $R_{500}$. In Fig.~\ref{fig:ebind_eagn} we show the profiles of non-gravitational and binding energy obtained from Eq. \ref{eq:Eng} and \ref{eq:Ebind}. We can see that the non-gravitational energy dominates over the binding energy out to $\sim0.3R_{500}$, such that the injected energy is sufficient to prevent the contraction of the inner regions and the formation of a cool core. 

Given the large required energy input, AGN activity is the most likely source of non-gravitational energy. Indeed, stellar feedback in the form of stellar winds and supernovae is known to be insufficient and too centrally concentrated to offset the cooling and regulate the star formation rate of massive galaxies \citep{Benson:2003,Kay:2003}. The age of the stellar populations of NGC~3298 implies that the galaxy has quenched some 10 Gyr ago. If the quenching was induced by a giant AGN outburst, the stellar age places the epoch of entropy injection around redshift $\sim$2-3. From the relation between black hole mass and velocity dispersion \citep{Kormendy:2013}, we estimate that the central SMBH should have a mass $M_\bullet \simeq 10^{9.5} M_\odot$. Assuming that the SMBH was accreting close to the Eddington rate and injecting a power of $\sim10^{45}$ erg/s into the system, the SMBH should have remained active for a total of $\sim1$ Gyr. Such prolonged episodes of AGN activity may have raised the entropy and the cooling time of the gas to the point where the gas could no longer cool down and condense, thereby preventing the formation of a classical cool core. This would explain the absence of clear AGN feedback features at present times, as evidenced by the low power and small spatial extent of the radio jets. While AGN outbursts of such scales have been observed in several cases \citep{McNamara:2005,Giacintucci:2020}, these were found in much more massive systems, such that the injected non-gravitational energy did not destroy the surrounding cool core.

\subsection{The compact core}

\begin{figure*}
\centerline{\includegraphics[width=0.5\textwidth]{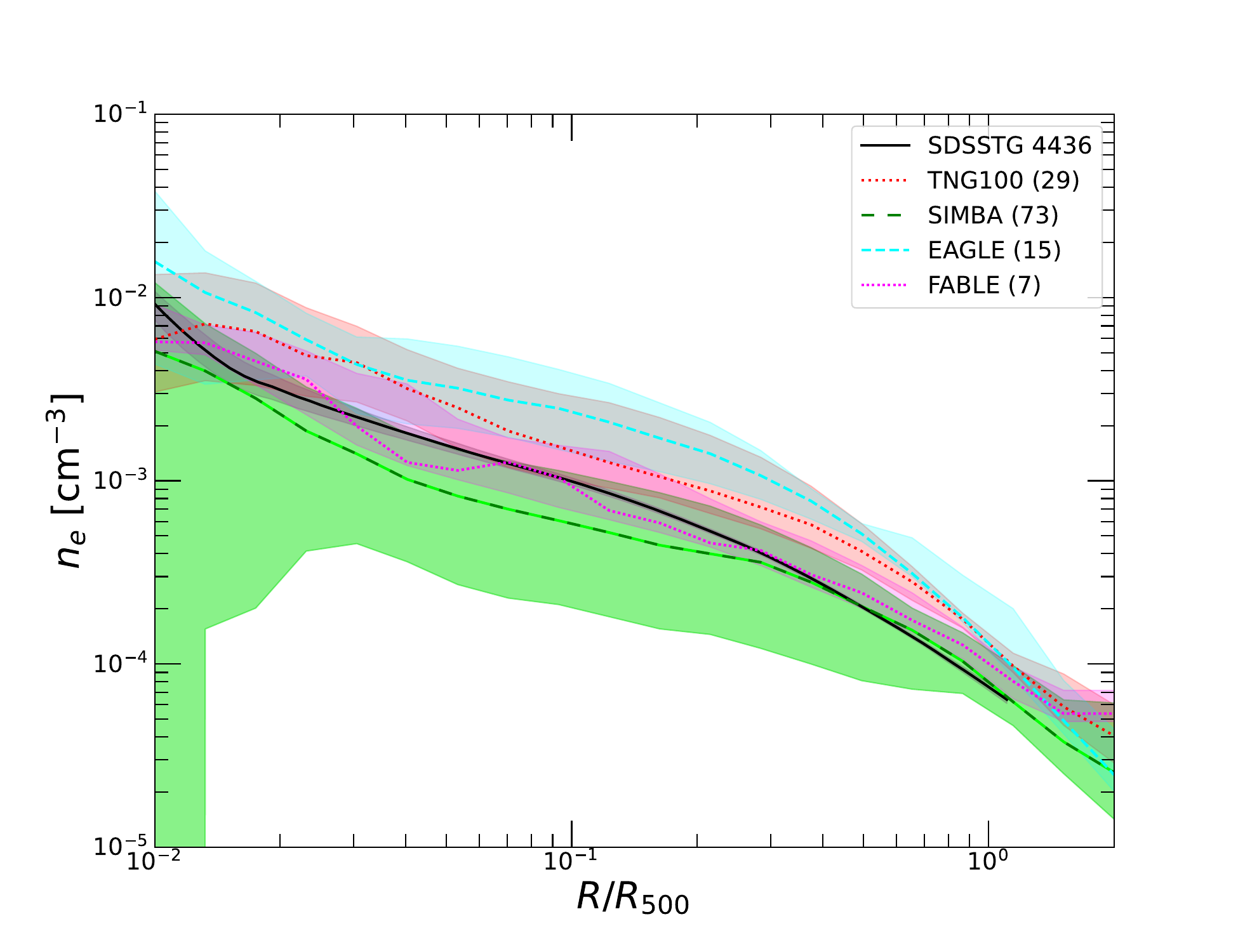}\includegraphics[width=0.5\textwidth]{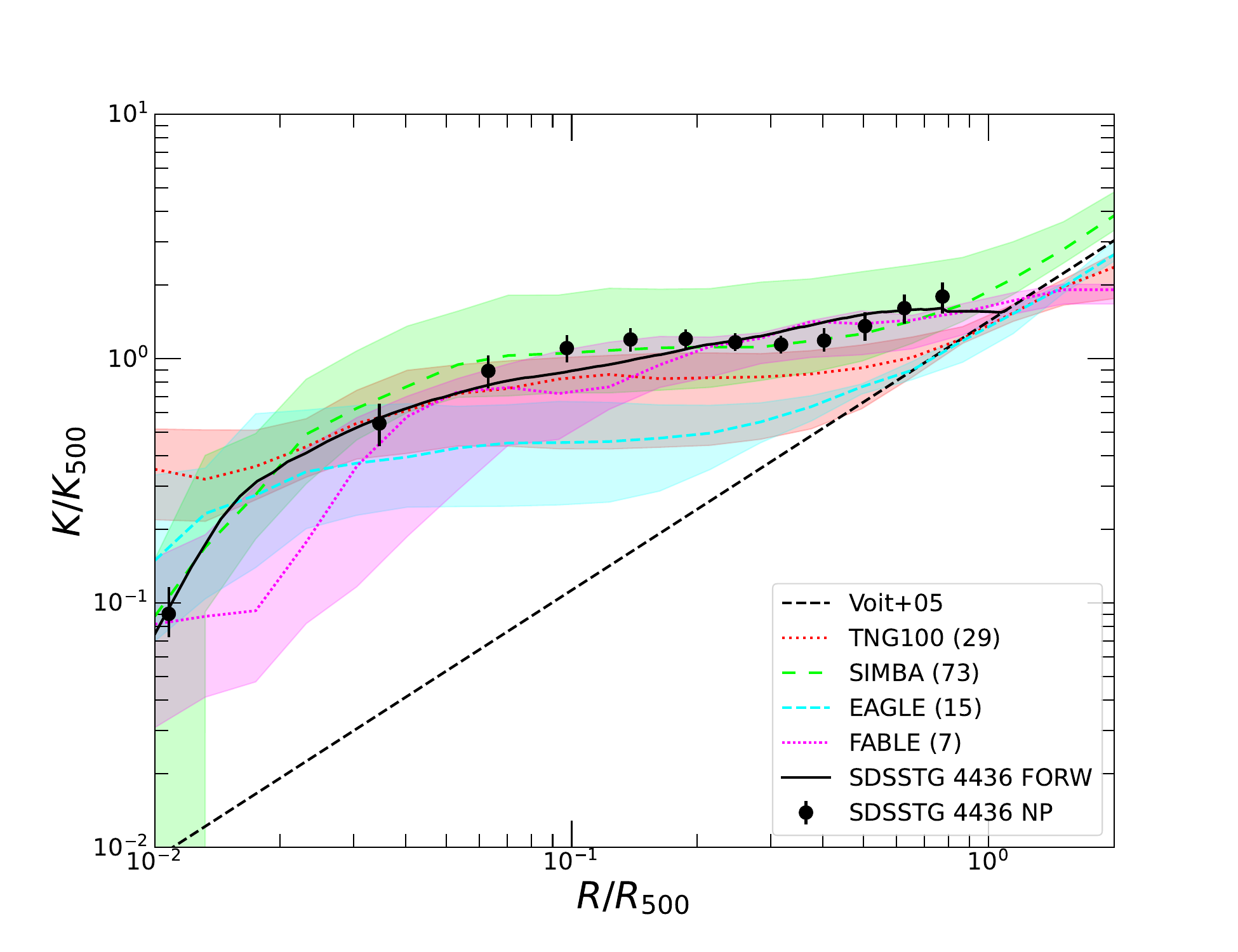}}
\caption{\label{fig:sim_median} Median and dispersion of gas thermodynamic profiles in four different simulations with various AGN feedback implementations. \textit{Left}: Electron number density profiles for galaxy groups in the TNG100 (dotted red), SIMBA (long dashed green), EAGLE (short dashed cyan), and FABLE (dashed magenta) simulations. In each case, the curve shows the median of the population, whereas the shaded area shows the $1-\sigma$ scatter around the median. The density profile of S4436 is indicated as the black curve. The numbers in parenthesis show the number of selected halos from each simulation set. \textit{Right}: Same as the left-hand panel for the self-similar scaled gas entropy.}
\end{figure*}

While the cooling time beyond $\sim0.03R_{500}$ exceeds the Hubble time, within the compact core the cooling time decreases steeply and the gas is efficiently cooling. The gas density profile of the system (Fig. \ref{fig:3dprofs}) clearly shows the system is made of two separate components, with the compact core being confined within the central galaxy. The compact core is unresolved by \emph{XMM-Newton} ($R<10$ kpc) and features a high metallicity. The associated gas mass is a small fraction of the stellar mass ($M_{\rm gas}(<10\mbox{ kpc})\sim10^{9}M_\odot$ compared to a total stellar mass of $\sim3\times10^{11}M_{\odot}$). Given its small size and high metallicity, the compact core resembles the ``coronae'' of elliptical galaxies in massive clusters \citep{Sun:2007,Liu:2024,Tumer:2019}. Its gas content may have been replenished over time by stellar mass loss \citep{Sun:2007,OSullivan:2011_4261}, which would explain the high metallicity of the gas in the core and the steep metallicity gradient. The cooling of the corona may be responsible for powering the currently observed radio jets, which are confined within the compact core. The system may have established a cooling-heating balance on small scales between the cooling gas of the corona and the low-power jets, whilst not injecting much energy at the present day into the surrounding IGrM. We note that such low-power compact jets are much more numerous than bright radio galaxies with very extended radio jets \citep{Sabater:2019}. Small-scale feedback loops similar to the case of NGC~3298  may thus be common within group-scale halos. 

\subsection{Comparison with numerical simulations}

To understand whether high-entropy systems like S4436 are expected to exist in state-of-the-art galaxy evolution simulations, we extracted thermodynamic profiles from halos in a similar mass range in four different simulation sets: TNG100 \citep{Weinberger:2018}, EAGLE \citep{Schaye:2015}, SIMBA \citep{Dave2019}, and FABLE \citep{Henden2018}. The feedback models implemented within these simulations vastly differ from one another, as some simulations include only thermal feedback (EAGLE) while others alternate between a thermal ``quasar mode'' at high accretion rate and a kinetic ``radio mode'' at low accretion rates. The implementation of the kinetic feedback can be either random (IllustrisTNG) or directional (SIMBA). For each simulation set, we selected halos in the mass range $4\times10^{13}M_\odot\leq M_{500}\leq 2\times10^{14}M_\odot$ and extracted their 3D gas density and entropy profiles. We scaled the profiles according to the true values of $R_{500}$ and $K_{500}$ as determined in the simulation. We then interpolated the profiles onto a logarithmically spaced common radial grid spanning the radial range $[0.01-2]R_{500}$, and calculated the median and dispersion of the self-similar scaled profiles. 

In Fig. \ref{fig:sim_median} we show the comparison between the median electron density and entropy profiles in the simulations and the deprojected S4436 profiles. On Fig. \ref{fig:sim_median} we can see that the electron density of S4436 lies below the typical predictions of EAGLE and TNG100, close to the median of FABLE and above the typical SIMBA profile. Similarly, the entropy profiles of TNG and EAGLE systems are usually lower than the value estimated here, especially at intermediate radii ($0.1-0.5R_{500}$). Conversely, in SIMBA the measured entropy is close to the profile reported here, and the upper boundary of the 1$\sigma$ envelope exceeds the observed profile by a factor of $\sim2$, which shows that much more extreme systems exist in the simulation. Looking at the distribution of the individual profiles in each simulation (see Fig. \ref{fig:sim_indiv}), we can see that the profile recovered for S4436 occupies the upper end of the TNG100 and FABLE entropy profile distributions, which is what we would expect if this system represents the extreme end of the population. Conversely, the electron density profiles of EAGLE halos all exceed the profile of S4436, which indicates that the feedback scheme implemented in EAGLE is too gentle to generate high-entropy systems like S4436. Finally, we can see that the bulk of the SIMBA halos feature a very high entropy around $0.1R_{500}$, with some systems being as much as three times more extreme than S4436. The strong feedback implemented in SIMBA was sufficient to evacuate these halos almost completely, which leads to extremely low densities across the entire volume. While observations of a single system are not sufficient to rule out this model, the rather exceptional nature of S4436 renders this scenario improbable. Indeed, the adopted feedback mechanism must be flexible enough to reproduce at the same time high-entropy systems like S4436 and classical X-ray selected groups such as NGC 5044 \citep{Gastaldello:2009,Schellenberger:2021}, which were able to retain their cool core until this day.

Comparing the shape of the simulated profiles with the data, we notice that TNG100 systems typically feature extended high-entropy cores, and thus cannot reproduce the sharp drop at small radii observed in S4436 because of the presence of the central compact core. Therefore, in TNG100 the AGN energy is injected close to the centre of the system, which prevents the formation of a high-density core. This behaviour is also present, to a lesser extent, in the EAGLE profiles. Conversely, FABLE and SIMBA profiles qualitatively reproduce the shape observed in S4436, and their entropy typically falls off steeply inside $\sim0.05R_{500}$. This likely implies that the bulk of the AGN energy in these simulations is injected outside of the central galaxy, which allows for the formation of compact low-entropy cores. 

To understand possible formation paths for systems like S4436, we identified two halos in FABLE that feature a very similar entropy profile as S4436 at $z=0$ and traced their evolution. We extracted their entropy profiles at 5 different redshifts from $z=0.8$ until today and studied the injection of entropy into these systems. We found that in both halos, the high-entropy core was not in place before $z=0.4$ and the entropy was raised quickly at $z<0.4$ by strong feedback episodes. This is induced by a switch from quasar mode at higher redshifts to the more efficient radio mode at later times in this simulation. If that is the case, traces of recent feedback episodes should still be found in the IGrM in the form of ancient buoyantly rising cavities, that may reach the virial radius of the system and possibly lead to gas ejection from the halo. Given the fast cooling rate of radio-emitting electrons, the absence of large-scale radio features only places a lower limit of a few hundred Myr on the epoch of the latest feedback episode. Deeper observations of this system with \emph{XMM-Newton} planned for the next observation cycle will test this scenario by searching for traces of previous feedback activity in this system.

\section{Conclusion}

In this paper, we reported on multi-wavelength observations of the galaxy group S4436 centred on the massive elliptical galaxy NGC 3298, which shows unusual thermodynamic properties. Our findings can be summarised as follows,

\begin{itemize}
\item The X-ray emission from the group is regular and highly centrally peaked, with the X-ray peak coinciding with the dominant galaxy NGC 3298. The large magnitude gap (2.1 mag in $r$ band) between the central galaxy and the second brightest member classifies the system as a fossil group. Altogether, this implies that the system is relaxed and has not experienced a merger in a long time. The mean temperature of $1.85\pm0.07$ keV and the velocity dispersion of $574\pm74$ km/s estimated from 31 spectroscopic members implies the system has a mass in the range $M_{500}\sim(6-10)\times10^{13}M_\odot$.

\item The entropy and cooling time profiles of the group rise steeply with radius beyond the dominant galaxy (Fig. \ref{fig:3dprofs}). The cooling time of the gas reaches the age of the Universe at 20 kpc ($\sim0.03R_{500}$) from the centre, where the measured entropy level ($\sim200$ keV cm$^2$) exceeds the gravitational collapse expectation by more than an order of magnitude. Given the relaxed dynamical state of the system, the entropy of the gas could not have been raised by a recent merging event, such that the entropy must be of non-gravitational origin.

\item High-resolution radio observations with LOFAR VLBI reveal the existence of low-power ($P_{\rm 144 MHz}\sim1.4\times10^{23}$ W/Hz), compact ($\sim5$ kpc) radio jets (Fig. \ref{fig:lofar_manga}). The radio jets are confined within the central galaxy and are not currently injecting energy into the surrounding IGrM. The central galaxy is massive ($\log (M_\star/M_\odot)\sim11.5$) and has a low star formation rate, with no detection of H$\alpha$ emission. Given the high stellar age ($\sim11$ Gyr) and the absence of radio emission beyond the central low-power radio lobes, the bulk of the observed high entropy must have been injected in the past.

\item Inside $R_{500}$, the total injected non-gravitational energy estimated from the excess heat with respect to the gravitational collapse expectation is $\sim1.5\times10^{61}$ erg (see Sect. \ref{sec:energy}), which is comparable to the total binding energy of the IGrM ($\sim4\times10^{61}$ erg). The non-gravitational energy dominates over the binding energy out to $\sim0.3R_{500}$, such that the excess heat is sufficient to unbind the gas particles within the group's core. Previous strong AGN outbursts may thus have raised the entropy level to the point that the gas could no longer condense and reform a cool core.

\item In the innermost regions, the system features a very compact ($<10$ kpc), dense core where the cooling time becomes short ($<1$ Gyr). The compact core resembles the coronae of elliptical galaxies in massive clusters \citep{Sun:2007}. The metallicity of the gas transitions from the Solar value in the compact core to $\sim0.2Z_\odot$ at 50 kpc, which shows that the compact core and the large-scale halo have a different origin. The gas mass within the compact core is less than 1\% of the stellar mass of the galaxy and may have been replenished by stellar mass loss.

\item Comparing the thermodynamic properties of S4436 with four different numerical simulation suites (TNG100, EAGLE, SIMBA, and FABLE), we found that the entropy profile of this system occupies the upper boundary of the entropy profiles in the TNG and FABLE simulations, which would be expected if this system represents the extreme of the group population. We do not find any comparable system within EAGLE, whereas similar systems are commonplace in SIMBA, which contains even more extreme objects. This probably implies that the implemented feedback is too gentle in EAGLE and too energetic in SIMBA. Comparison between simulations and observations over a representative sample of galaxy groups is needed to further constrain the feedback model implemented in these simulations.

\end{itemize}

\begin{acknowledgements} 
Based on observations obtained with XMM-Newton, an ESA science mission with instruments and contributions directly funded by ESA Member States and NASA. DE and RS acknowledge support from the Swiss National Science Foundation (SNSF) under grant agreement 200021\_212576. This research was supported by the Munich Institute for Astro-, Particle and BioPhysics (MIAPbP) which is funded by the Deutsche Forschungsgemeinschaft (DFG, German Research Foundation) under Germany´s Excellence Strategy – EXC-2094 – 390783311. MS acknowledges support from NASA grant 80NSSC23K0148 and NASA {\em Chandra} grant GO2-23079X. MAB acknowledges support from a UKRI Stephen Hawking Fellowship (EP/X04257X/1). The material is based upon work supported by NASA under award number 80GSFC21M0002. ET acknowledges funding from the HTM (grant TK202), ETAg (grant PRG1006) and the EU Horizon Europe (EXCOSM, grant No. 101159513). WC is supported by the Atracci\'{o}n de Talento Contract no. 2020-T1/TIC-19882 granted by the Comunidad de Madrid in Spain, and the science research grants were from the China Manned Space Project. He also thanks the Ministerio de Ciencia e Innovación (Spain) for financial support under Project grant PID2021-122603NB-C21 and HORIZON EUROPE Marie Sklodowska-Curie Actions for supporting the LACEGAL-III project with grant number 101086388. MB acknowledges support from the Next Generation EU funds within the National Recovery and Resilience Plan (PNRR), Mission 4 - Education and Research, Component 2 - From Research to Business (M4C2), Investment Line 3.1 - Strengthening and creation of Research Infrastructures, Project IR0000034 – “STILES - Strengthening the Italian Leadership in ELT and SKA”.
FG, LL, SE acknowledge the financial contribution from the contracts Prin-MUR 2022 supported by Next Generation EU (M4.C2.1.1, n.20227RNLY3 {\it The concordance cosmological model: stress-tests with galaxy clusters}), ASI-INAF Athena 2019-27-HH.0, ``Attivit\`a di Studio per la comunit\`a scientifica di Astrofisica delle Alte Energie e Fisica Astroparticellare'' (Accordo Attuativo ASI-INAF n. 2017-14-H.0),
and from the European Union’s Horizon 2020 Programme under the AHEAD2020 project (grant agreement n. 871158). BDO's contribution was supported by Chandra Grant TM2-23004X. LL acknowledges the financial contribution from the INAF grant 1.05.12.04.01. KK acknowledges funding support from the South African Radio Astronomy Observatory (SARAO) and the National Research Foundation (NRF).

LOFAR is the LOw Frequency ARray designed and constructed by ASTRON. It has observing, data processing, and data storage facilities in several countries, which are owned by various parties (each with their own funding sources), and are collectively operated by the ILT foundation under a joint scientific policy. The ILT resources have benefitted from the following recent major funding sources: CNRS-INSU, Observatoire de Paris and Université d’Orléans, France; BMBF,
MIWF-NRW, MPG, Germany; Science Foundation Ireland (SFI), Department of Business, Enterprise and Innovation (DBEI), Ireland; NWO, The Netherlands; The Science and Technology Facilities Council, UK; Ministry of Science and Higher Education, Poland; Istituto Nazionale di Astrofisica (INAF), Italy. This research made use of the Dutch national e-infrastructure with support of the SURF Cooperative (e-infra 180169) and the LOFAR e-infra group. This research made use of the LOFAR-IT computing infrastructure supported and operated by INAF, including the resources within the PLEIADI special "LOFAR" project by USC-C of INAF, and by the Physics Dept. of Turin University (under the agreement with Consorzio Interuniversitario per la Fisica Spaziale) at the C3S Supercomputing Centre, Italy. The Jülich LOFAR Long Term Archive and the German LOFAR network are both coordinated and operated by the Jülich Supercomputing Centre (JSC), and computing resources on the supercomputer JUWELS at JSC were provided by the Gauss Centre for Supercomputing e.V. (grant CHTB00) through the John von Neumann Institute for Computing (NIC). This research made use
of the University of Hertfordshire high-performance computing facility and the LOFAR-UK computing facility located at the University of Hertfordshire and supported by STFC [ST/P000096/1].
\end{acknowledgements}

\bibliographystyle{aa}
\bibliography{biblio}

\appendix

\section{Deprojection and PSF deconvolution}
\label{app:deproj}

The deprojected Forward and NP profiles are obtained in a forward modeling way by combining the three-dimensional profiles of gas density and temperature, projecting them along the line of sight and adjusting them onto the measured surface brightness and temperature profiles. This analysis makes use of the \texttt{pyproffit} \citep{Eckert:2020} and \texttt{hydromass} \citep{Eckert:2022} Python packages. Here we summarise the adopted methodology. For more details and a validation of the techniques on simulated data, we refer the reader to \citet{Eckert:2020,Eckert:2022}. 

\begin{figure*}
\centerline{\includegraphics[width=0.5\textwidth]{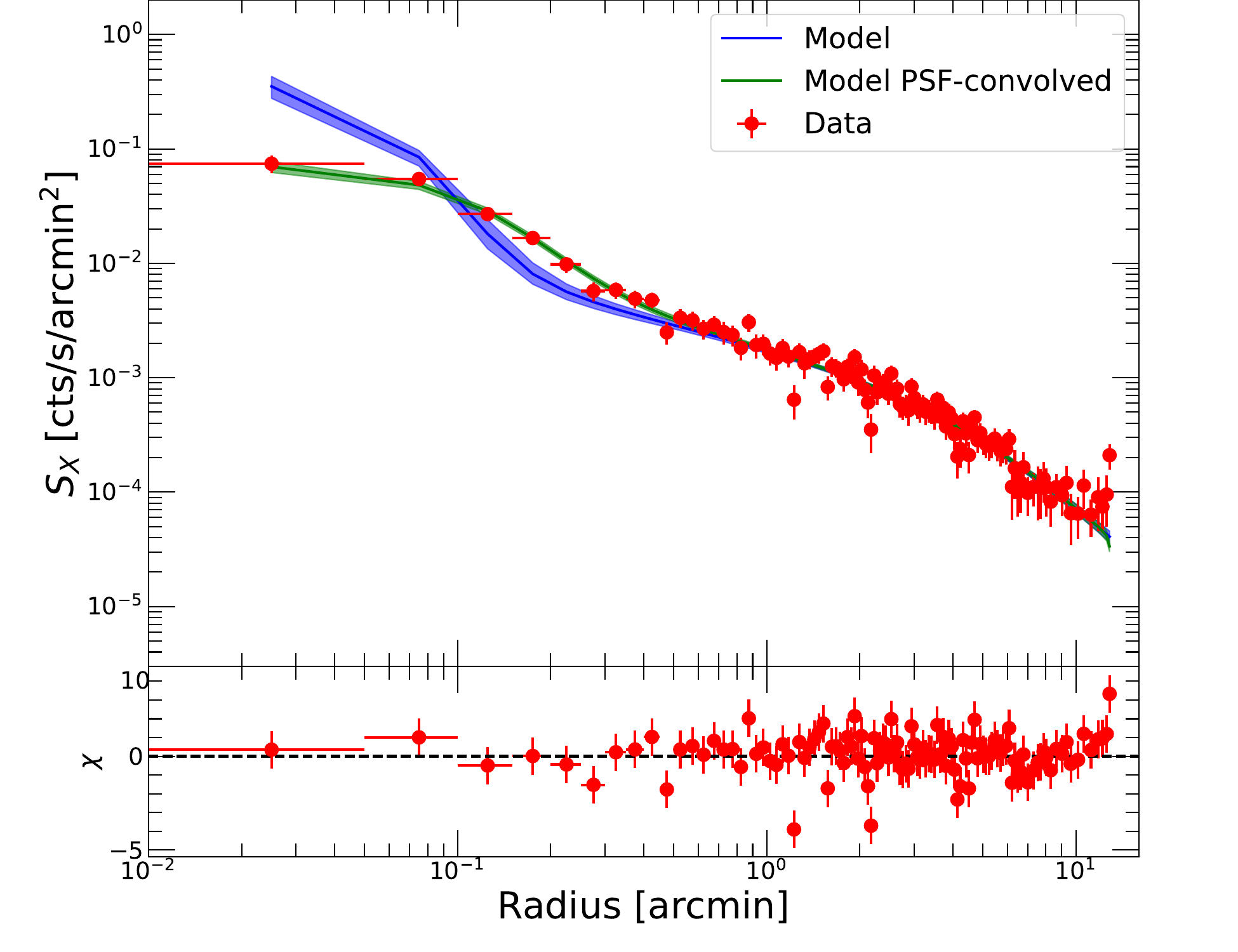}\includegraphics[width=0.5\textwidth]{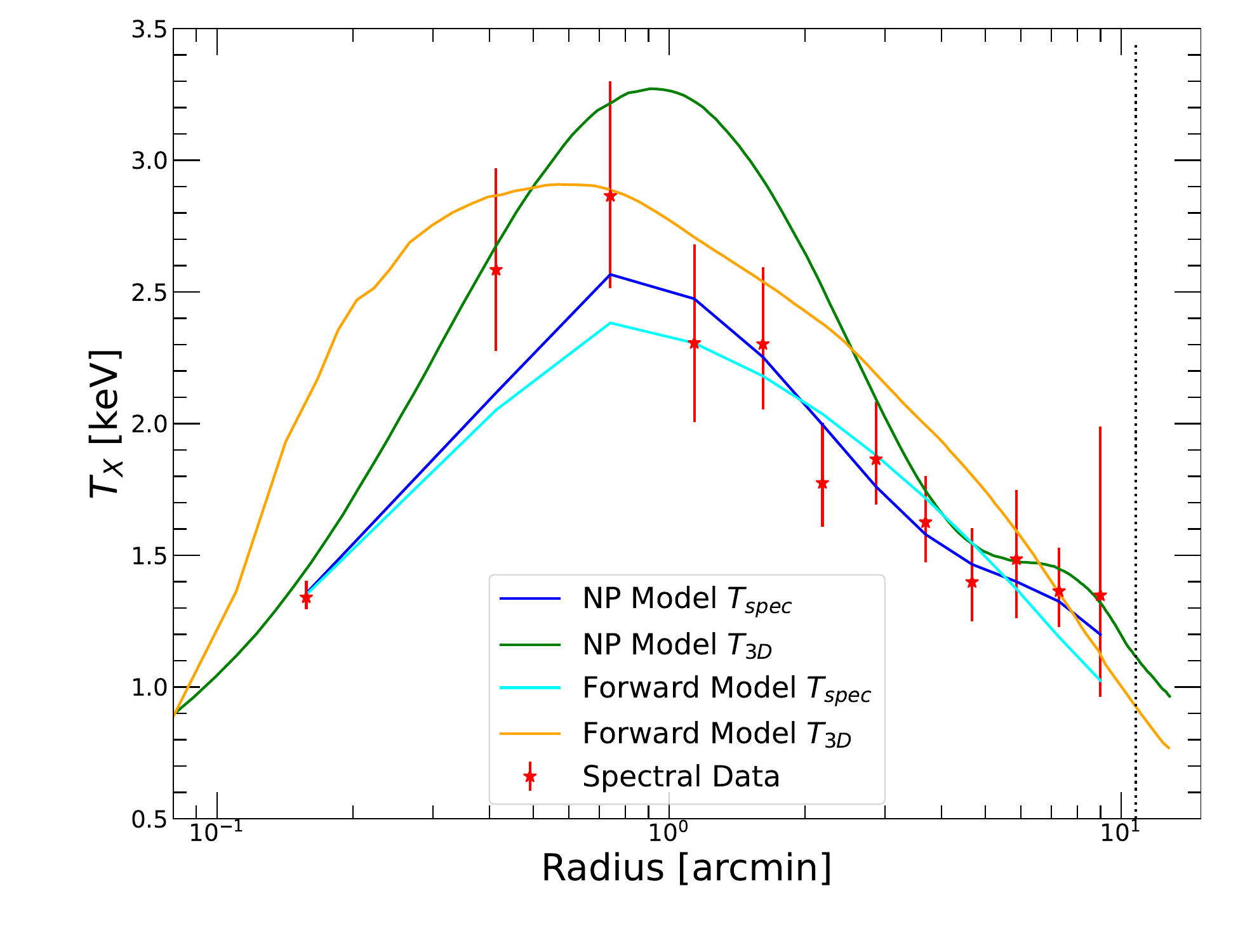}}
\caption{\label{fig:deproj} Deprojection and PSF deconvolution. \emph{Left:} Surface brightness profile in the [0.7-1.2] keV band (red points). The green curve shows the best fitting multi-scale model (green curve), projected along the line of sight and convolved with the \emph{XMM-Newton} PSF. The blue curve shows the model profile deconvolved from the PSF. In the bottom panel we show the residuals from the best fit model. \emph{Right:} Spectroscopic temperature profile (red points) and best fitting models. The orange and green curves show the 3D model profiles from the ``Forward'' and ``non-parametric'' methods, respectively. The blue and cyan curves show the corresponding projected and PSF convolved models.}
\end{figure*}

The gas density profile is estimated using the multi-scale deprojection technique introduced in \citet{Eckert:2020}. Specifically, the 3D emissivity profile $\epsilon(r)$ is described as a linear combination of radial basis functions,

\begin{equation}
\epsilon(r) = \sum_{p=1}^P \alpha_p \Phi_p(r).
\end{equation}

with $\{\Phi_p\}_{i=1}^P$ the adopted basis functions and $\{\alpha_p\}_{i=1}^{P}$ the associated coefficients. We adopt King functions as our choice of radial basis functions,

\begin{equation}
\Phi_p(r) = \left(1+\frac{r}{r_{c,p}} \right)^{-3\beta_p}
\end{equation}

with the parameters $r_{c,p}$ and $\beta_p$ governing the scale of each basis function and its outer slope, respectively. This choice is motivated by the fact that these functions describe a monotonous radial decline that is appropriate for galaxy clusters and groups, and can be analytically projected, such that the relation between the 2D and 3D profiles is analytic \citep{Eckert:2020}. For any given set of coefficients $\{\alpha_p\}_{p=1}^{P}$, the projected surface brightness profile can be predicted and fitted to the data. The choice of parameters $r_c$ and $\beta$ is set a priori to model monotonously declining profiles with a wide range of shapes across the radial range covered by the data. To model the \emph{XMM-Newton} PSF, we use the \citet{Read:2011} analytic model of the EPIC PSF to construct a PSF mixing matrix with the same binning as the surface brightness profile. The PSF mixing matrix is obtained by generating images of a flat surface brightness distribution within each annulus individually, and convolving the image in 2D with the \citet{Read:2011} model. The mixing matrix is then obtained by counting the fraction of the flux leaking into each surrounding annulus. The model surface brightness profile is then convolved with the mixing matrix to predict the observed surface brightness. The coefficients of the multi-scale model are then adjusted to reproduce the observed profile using the No U-Turn Sampler (NUTS) implemented within \texttt{PyMC} \citep{pymc3}. The resulting model fit is shown in the left-hand panel of Fig. \ref{fig:deproj}.

In parallel, we fit the spectroscopic temperature profile by projecting a model for the 3D temperature distribution along the line of sight. The projected model temperature is calculated as the emission-weighted mean of the line-of-sight temperature distribution. We describe here the two approaches considered in Sect. \ref{sec:deproj}. First, we apply a parametric reconstruction whereby the three-dimensional pressure profile is described as a generalized Navarro-Frenk-White profile \citep{Nagai:2007}, 
\begin{equation}P(r) = \frac{P_0}{ (c_{500}r)^{\gamma}  (1+(c_{500}r)^\alpha)^{(\beta-\gamma)/\alpha}}.\end{equation}
Since the parameters of the model are strongly degenerate, we fix the value of the middle slope, $\alpha$, to the value of 1.3 \citep{Arnaud:2010}. The four remaining parameters of the model ($P_0, c_{500}, \beta,$ and $\gamma$) are left free to vary. At each step, the pressure profile is combined with the density profile through the ideal gas equation to compute the model 3D temperature profile, which is then projected along the line of sight and convolved with the PSF mixing matrix to predict the spectroscopic temperature profile. We measure $P_0=(1.08\pm0.35)\times10^{-3}$ [keV cm$^{-3}$], $c_{500}=(2.22\pm0.59)\times10^{-3}$ [kpc$^{-1}$], $\beta=4.11\pm0.89$, and $\gamma=0.58\pm0.06$. The corresponding model is labelled as the ``Forward'' model and is shown as the cyan curve and shaded area in Fig. \ref{fig:3dprofs}. The fit of this model to the spectroscopic temperature profile is shown in the right-hand panel of Fig.~\ref{fig:deproj}.

For comparison, we also apply a non-parametric reconstruction in which the 3D temperature profile is described as a linear combination of log-normal functions \citep{Eckert:2022}. In this case, the model temperature profile is given by a combination of $N_g=200$ log-normal functions,
\begin{equation}
        T_{NP}(r)=\sum_{i=1}^{N_g} G_i \frac{1}{\sqrt{2\pi\sigma_i^2}}\exp\left(-\frac{(\ln(r)-\ln(\mu_i))^2}{2\sigma_i^2}\right)\label{eq:gp}
\end{equation}
with $\{\mu_i\}_{i=1}^{N_g}$ the mean radii of each function, that we choose to be logarithmically spaced from the centre to the outskirts. To smooth fluctuations on scales that are smaller than the adopted radial binning, the standard deviations $\{\sigma_i\}_{i=1}^{N_g}$ are set to the width of the nearest spectroscopic annuli. The model is then projected along the line of sight and convolved with the PSF. Tests of the method on mock data showed that it can accurately reproduce complex radial profiles of arbitrary shapes whilst at the same time smoothing out small-scale fluctuations. The blue data points in Fig. \ref{fig:3dprofs}, labelled ``NP'', are obtained by optimising the values of the coefficients $\{G_i\}_{i=1}^{N_g}$ and evaluating the best fitting 3D model at the middle radius of each spectroscopic bin. In the right-hand panel of Fig. \ref{fig:deproj} we show the fit of this model to the spectroscopic temperature data and the corresponding 3D model. Assuming that the gas is in hydrostatic equilibrium within the potential well set by the dark matter, we fit the resulting profiles with a Navarro-Frenk-White \citep[NFW,][]{Navarro:1996} density profile. This yields a mass $M_{500}=6.1_{-1.0}^{+1.8}\times10^{13}M_\odot$, which is slightly lower than, but consistent with, the value of $(7.8\pm1.6)\times10^{13}M_\odot$ obtained from the mass-temperature relation of \citet{Umetsu:2020}.

\section{Individual simulated thermodynamic profiles}

\begin{figure*}
\centerline{\includegraphics[width=0.5\textwidth]{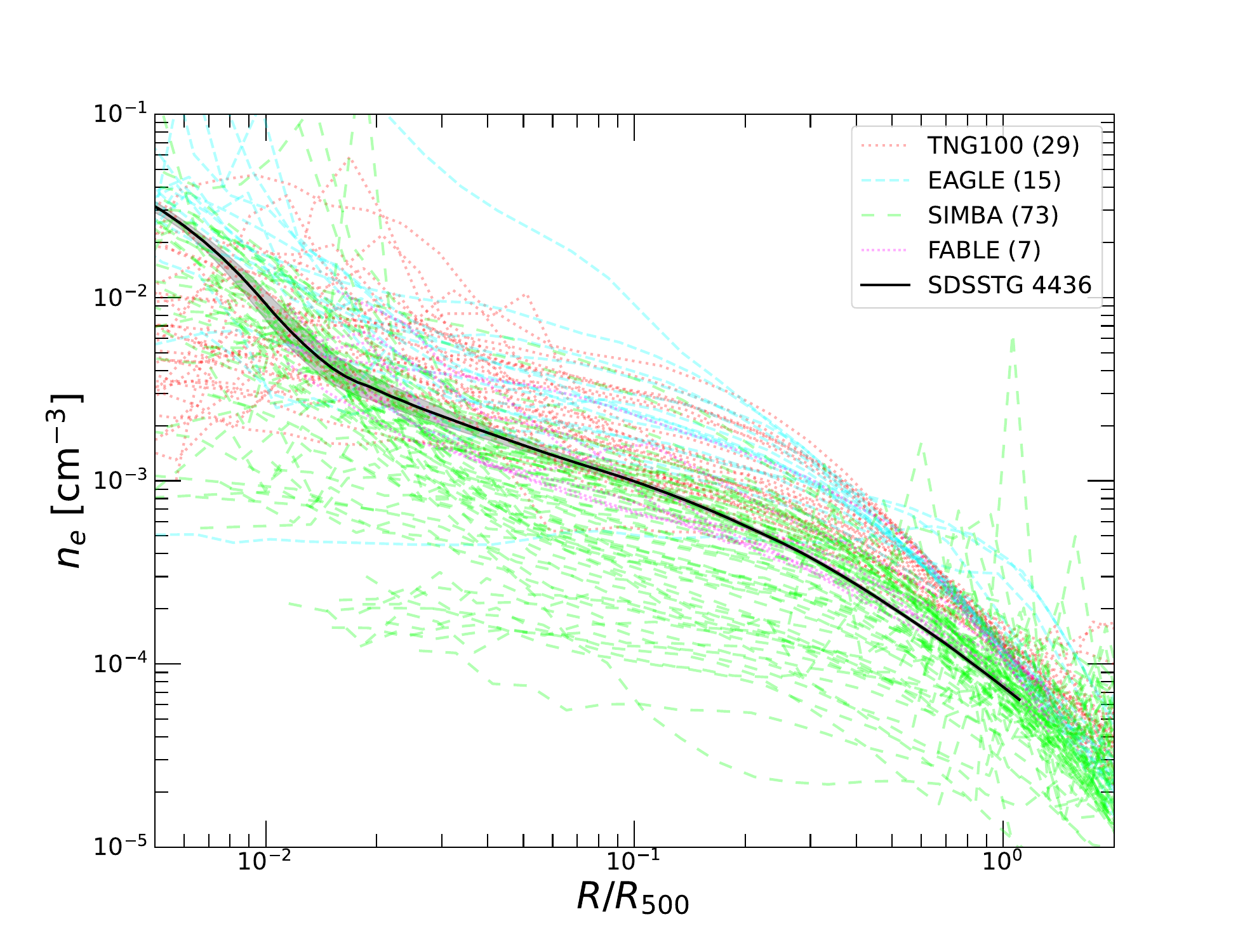}\includegraphics[width=0.5\textwidth]{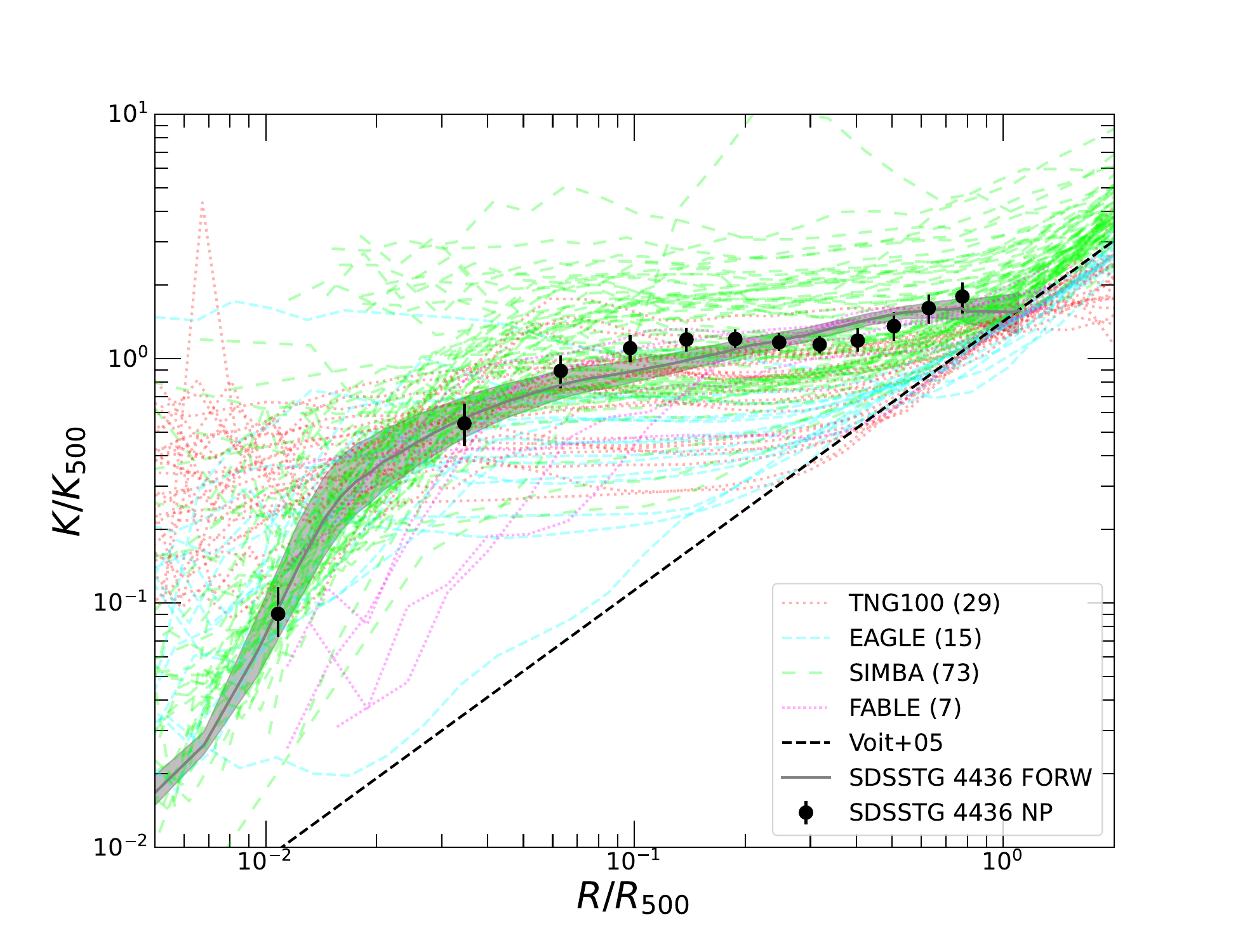}}
\caption{\label{fig:sim_indiv} \textit{Left}: Individual electron density profiles for halos in a similar mass range as S4436 extracted from TNG100 (dotted red), EAGLE (long-dashed cyan), SIMBA (short-dashed green), and FABLE (dotted magneta) simulations. The black curve and shaded area show the electron density profile of S4436 and its 1-$\sigma$ error envelope. The numbers in parenthesis show the number of halos considered here for each simulation set. \textit{Right}: Same as the left-hand panel for the self-similar scaled entropy profiles.}
\end{figure*}

In Fig. \ref{fig:sim_indiv} we show the individual electron density and entropy profiles used to generated Fig. \ref{fig:sim_median}. The profiles were extracted from halos in the mass range $4\times10^{13}M_\odot\leq M_{500}\leq 2\times10^{14}M_\odot$ from each simulation box. The thermodynamic profiles of S4436 are shown for comparison.

\end{document}